\documentclass[fleqn,usenatbib]{mnras}

\usepackage{newtxtext,newtxmath}

\usepackage[T1]{fontenc}
\usepackage{ae,aecompl}
\usepackage{tablefootnote}

\usepackage{graphicx}	
\usepackage{amsmath}	
\usepackage{amssymb}	

\usepackage{siunitx}
\usepackage{booktabs}
\usepackage{multirow}
\usepackage[export]{adjustbox}

\newcommand\water{H$\sb{2}$O}
\newcommand\methane{CH$\sb{4}$}
\newcommand\carbdiox{CO$\sb{2}$}
\newcommand\ammonia{NH$\sb{3}$}
\newcommand\molhyd{H$\sb{2}$}

\DeclareSymbolFont{UPM}{U}{eur}{m}{n}

\title[Gaussian Processes \& T$_{\rm eff}$]{Estimating dayside effective temperatures of hot Jupiters and associated uncertainties through Gaussian process regression}

\author[Pass et al.]{
Emily K. Pass,$^{1,2,3}$\thanks{E-mail: ekpass@edu.uwaterloo.ca}
Nicolas B. Cowan,$^{1,2,4,5}$
Patricio E. Cubillos$^{6}$
and Jack G. Sklar$^{4,7}$
\\
$^{1}$McGill Space Institute, 3550 rue University, Montreal, QC, H3A 2A7, Canada\\
$^{2}$Institut de Recherche sur les Exoplan\`etes, Universit\'e de Montreal, C.P. 6128, Succ. Centre-ville, Montreal, QC H3C 3J7, Canada\\
$^{3}$Department of Physics \& Astronomy, University of Waterloo, 200 University Ave W, Waterloo, ON, N2L 3G1, Canada\\
$^{4}$Department of Physics, McGill University, 3600 rue University, Montreal, QC, H3A 2T8, Canada\\
$^{5}$Department of Earth \& Planetary Sciences, McGill University, 3450 rue University, Montreal, QC, H3A 0E8, Canada\\
$^{6}$Space Research Institute, Austrian Academy of Sciences, Schmiedlstrasse 6, A-8042, Graz, Austria\\
$^{7}$School of Computer Science, McGill University, 3480 Rue University, Montreal, QC H3A 2A7, Canada
}

\date{Accepted XXX. Received YYY; in original form ZZZ}

\pubyear{2019}

\begin{document}
\label{firstpage}
\pagerange{\pageref{firstpage}--\pageref{lastpage}}
\maketitle

\begin{abstract}
In this work, we outline a new method for estimating dayside effective temperatures of exoplanets and associated uncertainties using Gaussian process (GP) regression.  By applying our method to simulated observations, we show that the GP method estimates uncertainty more robustly than other model-independent approaches.  We find that unbiased estimates of effective temperatures can be made using as few as three broad-band measurements (white-light \textit{HST} WFC3 and the two warm \textit{Spitzer} IRAC channels), although we caution that estimates made using only IRAC can be significantly biased.  We then apply our GP method to the twelve hot Jupiters in the literature whose secondary eclipse depths have been measured by WFC3 and IRAC channels 1 and 2:  \hbox{CoRoT-2 b}; \hbox{HAT-P-7 b}; \hbox{HD 189733 b}; \hbox{HD 209458 b}; \hbox{Kepler-13A b}; \hbox{TrES-3 b}; \hbox{WASP-4 b}; \hbox{WASP-12 b}; \hbox{WASP-18 b}; \hbox{WASP-33 b}; \hbox{WASP-43 b}; and \hbox{WASP-103 b}. We present model-independent dayside effective temperatures for these planets, with uncertainty estimates that range from $\pm$ 66 K to $\pm$ 136 K.
\end{abstract}


\section{Introduction}
\subsection{Effective temperature}
The effective temperature of an exoplanet---i.e., the temperature of a perfect blackbody emitting the same flux and with the same radius as the real planet---is an important quantity.  Together with surface gravity, it is one of the most basic parameters for modelling planetary atmospheres \citep[for review, see][]{Fortney_2018}.  In particular, effective temperature estimates allow for analyses of Bond albedo and day-to-night heat transport \citep[e.g.,][]{Cowan_2011, Crossfield_2012, Schwartz_2015}.

Information on the planet's dayside can be obtained through measurements of the secondary eclipse, where the planet is observed to pass behind its host star.  For an edge-on system, the secondary eclipse depth is related to the planet's spectral radiance by \citep{Cowan_2007}:
\begin{equation}
\label{eq:eclipse}
\frac{F_\textrm{p,day}}{F_*} = \frac{B_\lambda(T_\textrm{b,p})}{B_\lambda(T_{\textrm{b},*})}\Bigg(\frac{R_\textrm{p}}{R_*} \Bigg)^2 + A_{\textrm{g},\lambda} \Bigg(\frac{R_\textrm{p}}{a} \Bigg)^2,
\end{equation}
\noindent wherein $F_\textrm{p,day}/F_{*}$ is the secondary eclipse depth, $R_\textrm{p} / R_*$ is the ratio of planetary to stellar radii, and $B_\lambda(T_\textrm{b,p}) / B_\lambda(T_{\textrm{b},*})$ is the ratio of spectral radiances for a planet with brightness temperature $T_\textrm{b,p}$ around a host star with brightness temperature $T_{\textrm{b},*}$ at wavelength $\lambda$.  The second term, which depends on the wavelength-dependent geometric albedo $A_{g,\lambda}$ and the planet's semi-major axis $a$, is the contribution from reflected starlight. For sufficiently hot planets, reflected light can be neglected in the infrared regime, as the planet's thermal emission dominates. For this work, we therefore ignore the contribution of reflected light. However, we note that this is an approximation; for example, see \citet{Keating_2017}, who find reflected light is non-negligible for the hot Jupiter \hbox{WASP-43 b} at near-infrared wavelengths.

If secondary eclipse depths were measured at all wavelengths, the planet's dayside effective temperature could be calculated directly. By rearranging Equation~\ref{eq:eclipse}, secondary eclipse depth can be converted to planetary spectral radiance.  Integrating this spectral radiance over all wavelengths yields the total radiance, $I$, with effective temperature then determined using the Stefan-Boltzmann law:
\begin{equation}
\label{eq:teff}
T_{\rm eff} = \Big(\frac{\pi I}{\sigma}\Big)^{1/4}.
\end{equation}
\indent In reality, secondary eclipse depth is not known at all wavelengths. At best, a planet may have broad-band photometry from the Infrared Array Camera \citep[IRAC;][]{Fazio_2004} on \textit{Spitzer} \citep[\citealt{Werner_2004}; e.g.,][]{Deming_2007, Nymeyer_2011, Todorov_2014, Kammer_2015, Wong_2015} and some near-IR observations from ground-based telescopes \citep[e.g.,][]{Zhao_2012, Croll_2015, Zhou_2015, Martioli_2018}, as well as 1.1--\SI{1.7}{\micro\metre} spectral data from the Wide Field Camera 3 \citep[WFC3;][]{Cheng_2000} on the \textit{Hubble Space Telescope} \citep[\citealt{Bahcall_1986}; e.g.,][]{Crouzet_2014, Ranjan_2014, Line_2016, Cartier_2017, Mansfield_2018}.  More realistically, most planets are observed in only a small subset of these bands. In particular, planets with only 3.6 and \SI{4.5}{\micro\metre} measurements are common \citep[Figure 1 of][]{Schwartz_2015}, as these are the two IRAC channels which remain active on post-cryogenic \textit{Spitzer} \citep{Ingalls_2016}.

To determine the effective temperature, one must therefore interpolate/extrapolate the spectral radiance across all wavelengths, often from as few as two data points, or use spectral retrieval codes to generate spectra that are consistent with the observations. It is most insightful to convert spectral radiance to brightness temperature before interpolation, as---to first order---brightness temperature is constant as a function of wavelength. The inverse Planck function supplies the appropriate transformation:
\begin{equation}
\label{eq:bright}
T_\textrm{b,p}(\lambda) = \frac{hc}{\lambda k_\textrm{B}} \Bigg[\ln \Bigg(\frac{2hc^2}{\lambda^5 B_\lambda(T_\textrm{b,p})}\Bigg)+1\Bigg]^{-1}.
\end{equation}
\subsection{Estimation methods}
Generally, a planet's brightness temperature varies with wavelength as a result of the atmosphere's pressure-temperature profile and wavelength-dependent opacity.  However, blackbody-like behaviour is expected for isothermal, cloud-free atmospheres \citep{Seager_2005, Fortney_2006} or planets with high-altitude cloud decks \citep{Marley_1999}.  Bayesian Information Criterion \citep[BIC;][]{Schwarz_1978} analysis of exoplanets with secondary eclipse measurements in multiple bands has suggested that blackbody fits are often statistically favoured over detailed spectral fitting \citep{Hansen_2014, Garhart_2019}.  Such results motivate the Error-Weighted Mean (EWM) method of effective temperature estimation.  By assuming a Planck-like spectrum, the effective temperature can be calculated as simply the average of brightness temperature measurements, with measurements weighted by their respective uncertainties \citep{Schwartz_2015}.

Other model-independent estimates do not assume the planet behaves as a blackbody.  With the Linear Interpolation (LI) method, the brightness temperature is assumed to be constant at shorter wavelengths than the shortest wavelength observed and at longer wavelengths than the longest wavelength observed.  Between observations, brightness temperature is assumed to vary linearly with wavelength. This method implicitly gives more weight to measurements near the planet's blackbody peak and preserves any spectral detail in the observations \citep{Cowan_2011}, although it is not as robust against outliers as the EWM.

Model fitting with radiative transfer codes is another method for estimating dayside effective temperatures \citep[e.g.,][]{Fortney_2005, Stevenson_2010, Mancini_2013, Morley_2017}, with a variety of spectral retrieval frameworks currently in use (\citealt{Barman_2005}; \citealt{Irwin_2008}; \citealt{Line_2013}; \citealt{Benneke_2015}; \citealt{Waldman_2015}; \citealt{Lavie_2017}; \citealt{Henderson_2017}; \citealt{Gandhi_2018}; see review in \citealt{Madhusudhan_2018}). Such fitting algorithms are advantageous as they allow system parameters and physical constraints to further inform effective temperature estimates. Nevertheless, these estimates must be used with some caution. Most retrieval models have a dozen parameters and hence the fitting is underconstrained, even for the most observed planets. Outlier measurements can therefore have a dramatic impact on the resulting fit, leading to inaccurately-constrained parameters \citep{Hansen_2014}. Model-dependent methods are also only as accurate as the physics and assumptions included in the model; for example, \citet{Feng_2016} and \citet{Blecic_2017} show how the standard assumption of a single 1D thermal profile can bias the interpretation of a hot Jupiter's emission spectrum, while \citet{Line_2016} discuss the problematic assumption of a cloud-free atmosphere, as clouds strongly influence the spectra of brown dwarfs at similar temperatures. Even when clouds are taken into account, rigorous modelling of their effects is difficult due to the complexity of the physical processes involved in cloud formation \citep[e.g.,][]{Lee_2016, Lee_2017, Roman_2018}.  Model fitting also requires significant computational power, although retrieval frameworks powered by machine-learning algorithms aim to mitigate this issue \citep{MarquezNeila_2018, Zingales_2018}.

Model-independent effective temperature estimates can also be enhanced by machine learning.  In this work, we benchmark the performance of one such technique: Gaussian process (GP) regression. GPs introduce a Bayesian approach to effective temperature estimation, robustly evaluating uncertainties and yielding blackbody-like results when there is insufficient data to support a more complex fit.

In Section~\ref{sec:methods-gp}, we describe our GP method and motivate the selection of hyperparameters.  Our simulated observations are outlined in Section~\ref{sec:methods-sim}, with a description of our evaluation metric and alternate methods in Section~\ref{sec:methods-eval}. The performance of each method is benchmarked on simulated observations in Section~\ref{sec:res-sim} and applied to archival data in Section~\ref{sec:res-dat}. Results are summarized in Section~\ref{sec:summary}.

\section{Methods}
\label{sec:methods}

\subsection{Gaussian process regression}
\label{sec:methods-gp}

\subsubsection{Introduction to Gaussian process regression}

In this section, we provide a brief, intuitive description of GP regression. For a detailed, mathematical introduction to the technique, see \citet{Rasmussen_2006}.

GP regression provides a flexible approach to model fitting that offers distinct advantages over a traditional parameterized model. A parameterized model limits possible fits to a single class of functions; for example, you might assume the data follow a linear trend, and therefore search for the best-fitting parameters of the form $y=mx+b$. The EWM and LI methods operate in this way, with the former using a line of zero slope and the latter using piecewise-defined lines. In contrast, the GP method does not assume a specific functional form. A GP is therefore capable of fitting a linear trend, a quadratic trend, a sinusoidal trend, or other behaviours that are more complex and not known \emph{a priori}.

To do this, GPs require the specification of a covariance function (also called a kernel) and a mean function; a GP is completely specified by these two functions. The covariance function, perhaps unsurprisingly, specifies the covariance between points, and the parameters of this function are called hyperparameters. One of the simplest covariance functions is the squared exponential kernel, which takes the form:
\begin{equation}
k(r) = \sigma^2 \exp \Bigg( - \frac{r^2}{2l^2} \Bigg).
\end{equation}
The squared exponential kernel corresponds to an infinite set of basis functions, and it returns a covariance that depends only on the distance in input-space between points, $r$. This kernel is governed by two hyperparameters: $l^2$, the scale length, and $\sigma^2$, the signal variance. The scale length can be understood as the characteristic distance in input-space over which the output function varies, while the signal variance represents the characteristic amplitude of features. In this work, $r$ is the difference in frequency between two data points, $l$ is the characteristic value of $r$ over which features in the spectrum are correlated, and $\sigma$ is the typical amplitude of features in normalized brightness temperature (at the data's spectral resolution). Various methods may be used to set these hyperparameters, such as motivating them from physical constraints or optimizing them using the input data \citep{Ambikasaran_2015}. We adopt a hybrid modelling approach that borrows elements from both methods; this approach is the focus of Section~\ref{sec:gp-imp}.

A zero mean function is commonly used for GPs, although more complex functions can be chosen. In this work, we adopt mean functions that are constant but non-zero (namely, the error-weighted mean brightness temperature). The chosen mean function is most important in sparsely-populated regions of input-space, as our GP will take the value of this function in the absence of other data (i.e., when the nearest data point is many length scales away).

In summary, a GP is a covariance function and a mean function. Once these properties have been specified, including the hyperparameters of the covariance function, the GP can accept input data (with uncertainties) and output a fit to the data (with uncertainties). This output represents the range of functions consistent with the input data and covariance kernel. The GP's specificity results from an appropriate choice of hyperparameters for a given data set (which we will discuss in Section~\ref{sec:gp-imp}). However, it is important to note that even with specified hyperparameters, a GP still allows for much more flexible fits than could be achieved with a parameterized model (such as with the EWM or LI).

GPs have gained traction in the astrophysical community in recent years, being employed to fit stellar oscillations \citep{Brewer_2009}, model the occurrence rate density of exoplanets \citep{Foreman-Mackey_2014}, and estimate photometric redshifts \citep{Almosallam_2016}, among other applications. In the field of exoplanetary atmospheres, GPs have been used to account for instrument systematics in primary transit and secondary eclipse light curves \citep{Gibson_2012, Evans_2015}, although further applications remain to be explored.

These GP regressions primarily involve intermediate-sized data sets on the order of a hundred to a thousand data points ($N$).  Larger data sets have historically been intractable due to computation time scaling as $N^3$, although newer methods aim to improve performance for big data \citep{Foreman-Mackey_2017}.  However, our concern is with the opposite extreme: small data sets in which trends are sparsely sampled, often with as few as two or three measurements.  As these observations do not critically sample the underlying function, optimizing the hyperparameters from the observations themselves would result in inaccurate fits with poorly-estimated uncertainties.  However, this does not mean that GP regression is unusable in the sparse-data regime.  By physically motivating the hyperparameters using related data sets, appropriate hyperparameters can be chosen and information content retrieved, even with few observations.  

\subsubsection{Implementation of Gaussian process regression}

\label{sec:gp-imp}

We use the George library \citep{Ambikasaran_2015} for Python-based Gaussian process regression. Our GP uses a squared exponential kernel with the mean function set to the error-weighted mean. It therefore mimics the EWM method at wavelengths where the nearest observation is many length scales away. However, the EWM method assumes Gaussian-distributed uncertainties and involves weighting by the inverse of the variance ($w_i=1/\sigma_i^2$). This makes the method non-robust against outliers. Given the sparsity of our data, outliers can have a dramatic effect on our results; in our mean function, we therefore elect to weigh the points by the inverse of the error ($w_i=1/\sigma_i$). This weighting assumes the uncertainties follow a doubled-sided exponential distribution, although it is used more generally in situations where the uncertainties are mostly Gaussian but the presence of outliers is suspected \citep{Aster_2013}. In light of the above discussion, it is unsurprising that we find the $1/\sigma_i$ weighting outperforms the standard EWM method.

Before processing with the GP, the brightness temperature spectrum is normalized and converted from wavelength to frequency, as a linear length scale is expected in the frequency domain \citep{Rybicki_2004}.  We also convert these frequencies to petahertz to reduce the risk of overflow errors during data processing.

We consider two approaches to set the GP's hyperparameters. The first involves optimizing the two hyperparameters using SciPy's minimize routine \citep{Jones_2001} to minimize the negative likelihood of the GP model.  This approach is problematic for the sparse data that is currently available, as our sampling rate is much smaller than the expected length scale. This results in fits biased towards slowly-varying functions, as we have insufficient information to identify the covariance of our data.

Our second approach resolves this sampling problem by optimizing the length scale and signal variance using external data sets, where we do have enough information to constrain covariance. We must therefore identify an appropriate data set for estimating the characteristic length scale of hot Jupiter emission spectra.  While we expect structure over a variety of length scales, water vapour is the dominant opacity source at all relevant temperatures in the infrared \citep[Figure 6 of][]{Fortney_2018}.  For this reason, we estimate the length scale hyperparameter using the HITEMP \citep{Rothman_2010} line list database for H$_2$O.

\begin{figure}
\includegraphics[width=\columnwidth]{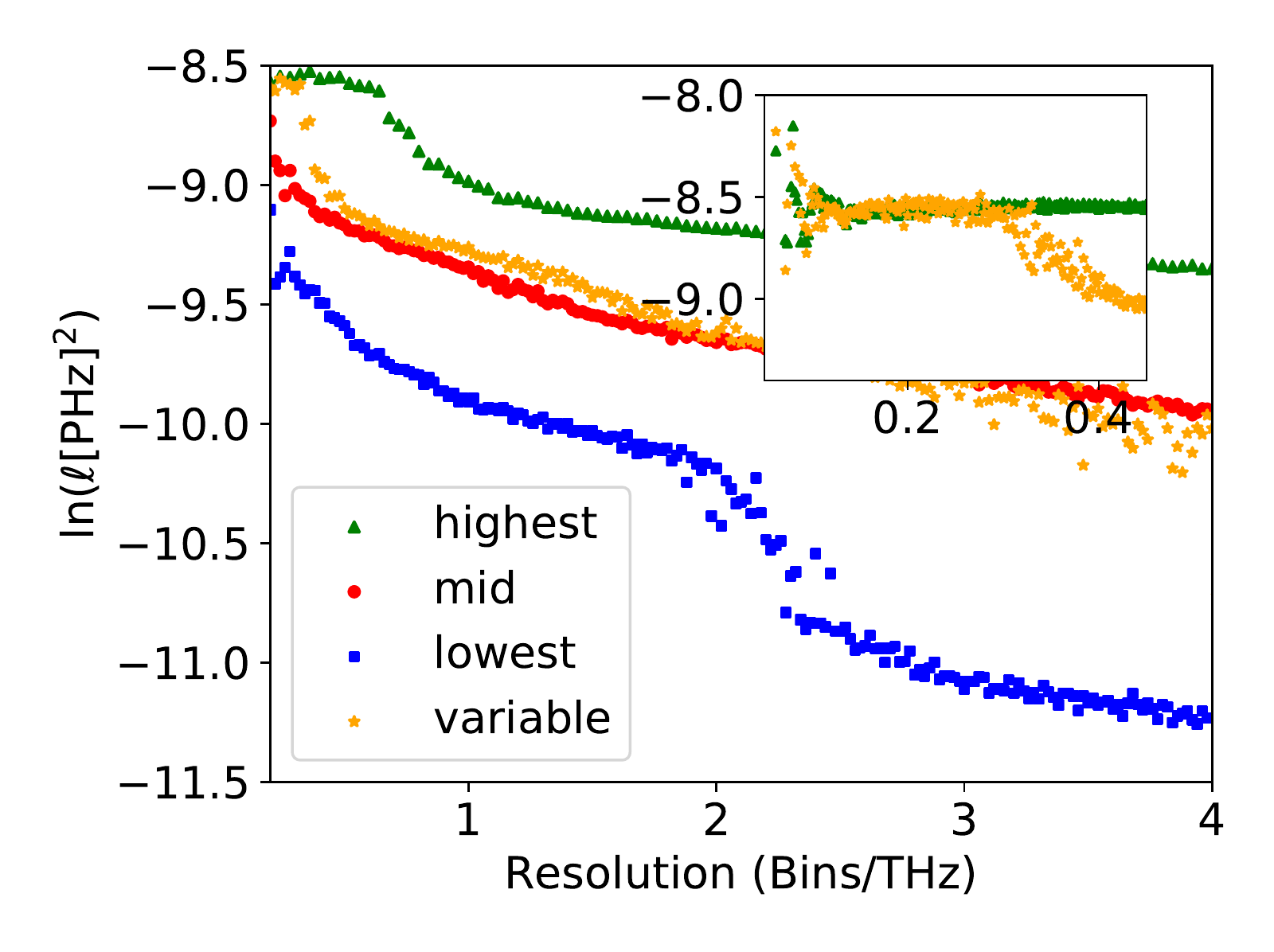}
\vspace{-0.5cm}
\caption{The maximum-likelihood log-length scale, ln($l^2$), is shown as a function of the number of bins in the water spectrum. As resolution increases, more small-scale structure is available to the GP and shorter length scales are favoured.  The different lines represent the uncertainty assumed, with the GP able to discount small-scale structure as random fluctuations given sufficient uncertainty. The lowest, mid, and highest lines show fixed, arbitrarily-selected uncertainty levels, while the variable line represents a dynamic uncertainty estimate based on the standard deviation of the binned spectral region.}
\label{fig:water_length_scale}
\end{figure}

Using the first approach, wherein we use SciPy's minimization routine to estimate hyperparameters, we analyze the intensity (wavenumber per column density) spectrum of water up to 5$\cdot 10^{14}$ Hz (\SI{0.6}{\micro\metre}). By varying the binning and noise for this spectrum (Figure~\ref{fig:water_length_scale}), we identify three significant length scales: the first at -8.55, the second between -10 and -9, and the third between -11.5 and -11. These length scales appear as plateaus or shallow downwards trends in Figure~\ref{fig:water_length_scale}; outside of these regions, length scale varies rapidly as the resolution of the spectrum is increased. Note that in these calculations, $l$ has been divided by units of PHz, resulting in unitless values for the log-length scale hyperparameter, ln($l^2$).

\begin{figure}
\includegraphics[width=\columnwidth]{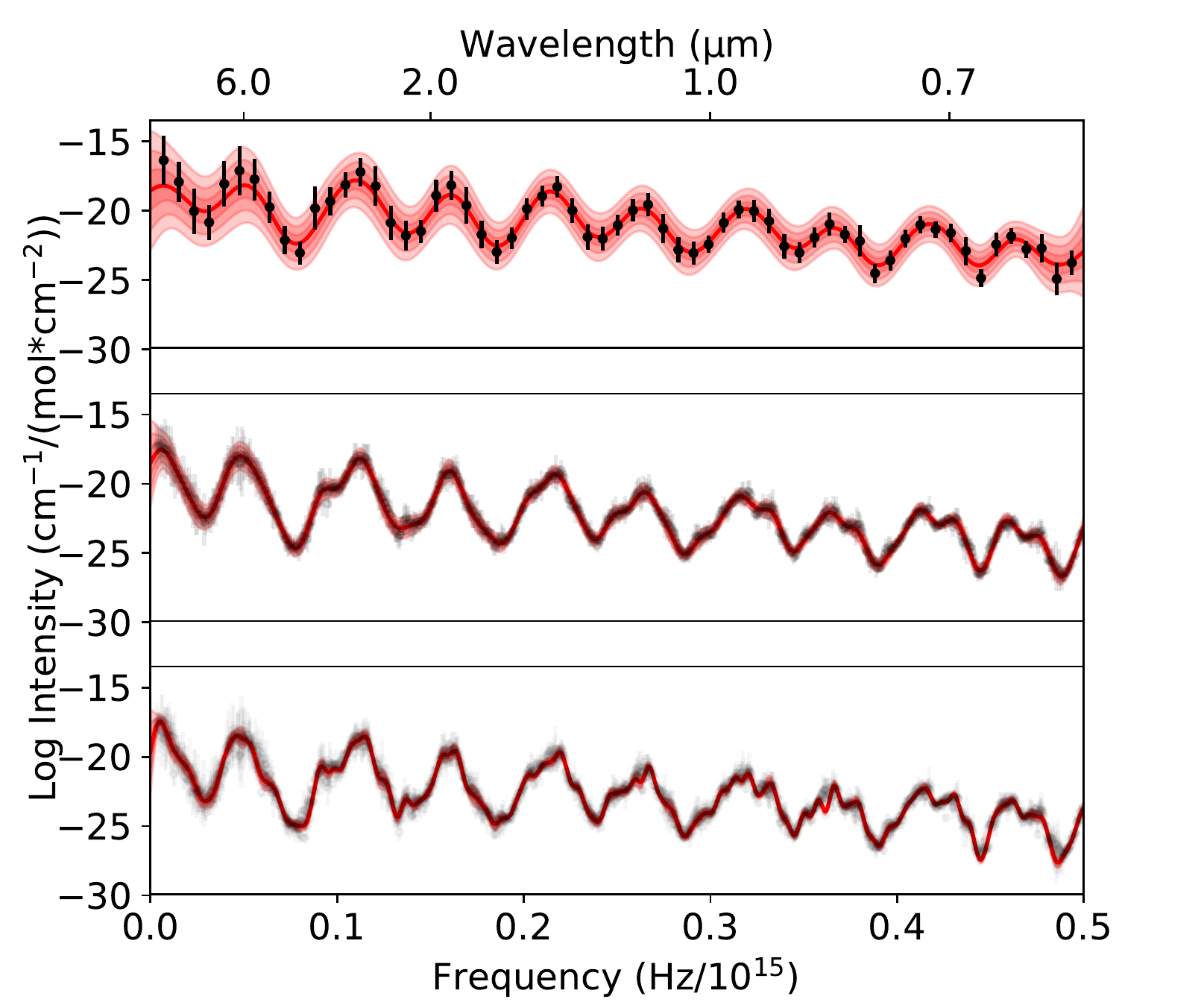}
\vspace{-0.2cm}
\caption{GP regression fits to the spectrum of water for low (top), mid (middle), and high (bottom) resolution binning. The red-shaded region indicates the 3$\sigma$ confidence interval.}
\label{fig:water_fits}
\end{figure}

We therefore find that water possesses structure over a variety of length scales. However, not all these scales are relevant for our purposes due to the finite spectral resolution and signal-to-noise of current observations. To identify which length scales correspond to relevant spectral details, we examine the GP fits to the water spectrum. We consider log-length scale hyperparameters of -8.55 (low-resolution binning), -9.62 (mid-resolution binning), and \hbox{-11.1} (high-resolution binning), and display the corresponding GP output in Figure~\ref{fig:water_fits}. Low-resolution binning allows for detection of the smooth periodic trend. Mid-resolution binning allows double peaks to be distinguished, while high-resolution binning reveals further fine structure. Given our desired precision and the quality of our brightness temperature data, only the smooth trend is relevant for our Gaussian process.  We therefore constrain the log-length scale hyperparameter to \hbox{$-8.55$ $\pm$ 0.05} (Figure~\ref{fig:water_length_scale}, inset), equivalent to $l=1.4\cdot10^{13}$ Hz, or \SI{0.19}{\micro\metre} at a wavelength of \SI{2}{\micro\metre}.  As we find that varying this hyperparameter within its uncertainty range does not perceptibly influence the results of our Gaussian process regression, we fix the log-length scale hyperparameter as $-8.55$.

While this investigation also yields signal variance hyperparameters, such estimates are not physically relevant; the y-axis of the water spectrum represents intensity (a molecular cross-section), which cannot be converted to brightness temperature without additional information about the planet's atmosphere (namely, the vertical temperature profile and the presence of clouds). A different data set is therefore required to inform our estimate of signal variance. This hyperparameter governs the amplitude of variations, thereby setting the uncertainty in the brightness temperature at frequencies far from observed data points.  Model emission spectra could be used for this task, although this would require setting a parameter space to explore, determining the physics to include in the models, estimating the relative abundance of planets with each composition, and so on. In short, such training would compromise the model-independent nature of GP regression.

As an alternative, we train the signal variance hyperparameter on the 54 hot Jupiters that have been observed in secondary eclipse, as listed in the exoplanets.org database \citep[][accessed July 2018]{Han_2014}.  After excluding planets that have been observed in only one band, planets lacking stellar parameters, and two planets with potentially erroneous database entries,\footnote{This latter category contains XO-3 b and HAT-P-2 b, for which the database lists an unrealistic \SI{4.5}{\micro\metre} secondary eclipse depth and uncertainty in \SI{8.0}{\micro\metre} secondary eclipse depth, respectively.} 37 hot Jupiters remain in the training set.  Using Equations~\ref{eq:eclipse} and~\ref{eq:bright}, we calculate brightness temperature spectra for these planets. For each star, stellar spectral radiance is determined using the Kurucz model \citep{Kurucz_1970, Castelli_2004} that best matches the star's effective temperature and surface gravity. Uncertainties in brightness temperature are propagated from uncertainties in known parameters using a 1000-iteration Monte Carlo. To consider the different fits that are possible as a result of these uncertainties, we use a random Gaussian distribution to generate 1000 versions of the observations and fit each set of observations using GP regression, setting the log-length scale to $-8.55$ and optimizing the signal variance with SciPy.

This method yields an ensemble of 37,000 signal variance hyperparameters centred at $-4$; as with the length scale, this is the natural logarithm of the $\sigma^2$ of Section~\ref{sec:methods-gp}. As our brightness temperature spectra are normalized, this value can easily be converted to a percentage amplitude using  \hbox{ $\sqrt[]{e^{-4}}$ = 0.14,} or 14\%. In other words, at the low spectral resolution of currently-available observations, we expect $\sim$14\% changes in brightness temperature as a function of wavelength. This is broadly in agreement with our expectations from atmospheric physics: the skin layer of the atmosphere has temperature $T_{\rm skin} = (1/2)^{1/4}T_{\rm eff}$ \citep{Pierrehumbert_2010}, corresponding to 16\% deviations from the planet's effective temperature. As the skin temperature represents the atmosphere's coolest layer, this suggests that 16\% is a reasonable upper limit on the potential variability. The presence of clouds would decrease the amplitude of the variations.

In general, one should set the GP's signal variance through a random sampling of the 37,000 estimated hyperparameters. However, the training data are not ideal. As the uncertainties in the training data are large, the upper tail of the hyperparameter distribution allows for highly inflated uncertainties in our GP-determined temperatures. For this reason, we fix the log-signal variance hyperparameter as $-4$, or 14\%.  This choice is consistent with theoretical expectations, as we have discussed. It also becomes strongly motivated following our analysis in Section~\ref{sec:res-sim}: with this hyperparameter, we retrieve statistically-appropriate distributions of effective temperature estimates.

\subsection{Simulated data set}
\label{sec:methods-sim}

\subsubsection{Model spectra}

To benchmark the performance of the Gaussian process method and compare its efficacy to other approaches, we require a suite of simulated emission spectra from which to generate sample observations. We create this suite with the open-source Python Radiative Transfer in a Bayesian framework (Pyrat Bay\footnote{\href{http://pcubillos.github.io/pyratbay}{http://pcubillos.github.io/pyratbay}}; Cubillos et al., in prep.). Six theoretical emission spectra are computed for each of the 54 hot Jupiters that have been observed in secondary eclipse, as listed in the exoplanets.org database as of July 2018 \citep{Han_2014}. The resulting sample of 324 model emission spectra provides a diverse set of hot Jupiter atmospheres with known effective temperatures on which to test our measurement technique.

Pyrat Bay, which is based on the Bayesian Atmospheric Radiative Transfer package \citep{Blecic2016phdThesis, Cubillos2016phdThesis}, produces 1D atmospheric models and combines them with a line-by-line radiative transfer module to model the emission spectra over the planet's dayside hemisphere.  We assume cloud-free atmospheres in hydrostatic, thermochemical, local-thermodynamic, and radiative equilibrium.  The atmospheric domain ranges from $10^{-8}$ to $10^{2}$ bar, sampled with 120 layers equidistant in logarithmic space. For the atmospheric composition, we adopt models with scaled solar metallicities; i.e., we start with solar elemental mixing ratios \citep{AsplundEtal2009araSolarComposition}, scaling the abundances of all elements beyond H and He.  Given the temperature, pressure, and elemental compositions, we compute the thermochemical equilibrium abundances using the open-source Thermochemical Equilibrium Abundances (TEA) code \citep{BlecicEtal2016apsjTEA}.

To attain radiative equilibrium, we follow the iterative procedure described in \citet{MalikEtal2017ajHELIOS}; i.e., we compute the upward and downward fluxes across the atmosphere under the two-stream approximation (see their Equation 10, with $\epsilon=2.0$).  On each iteration, the code updates the temperature profile such that the divergence of the bolometric net fluxes converges to a negligible value at each layer.  During this process, since the thermochemical equilibrium calculation is the most computationally demanding task, we update the chemistry only every eight iterations.

The domain of the spectrum ranges from 0.3 to \SI{33}{\micro\metre}, sufficient to contain the bulk of the stellar and planetary radiation.  At the top of the atmosphere, we model the incident stellar irradiation as a blackbody function at the effective temperature of the star, attenuated by a Bond albedo $A_\textrm{B}$, and considering good day--night energy redistribution over the planet.  At the bottom of the atmosphere, we include an internal radiative heat term corresponding to a 100~K blackbody. This choice is commonly made elsewhere in the literature \citep[e.g.,][]{Fortney_2007}, as it is approximately the value for Jupiter. However, this term has a negligible impact on the results.

We consider opacities from the main spectroscopically active species expected for exoplanets at these wavelengths: {\water} and {\carbdiox} from \citet{Rothman_2010}; {\methane}, {\ammonia}, and HCN from \citet{YurchenkoTennyson2014mnrasExomolCH4}; CO from \citet{LiEtal2015apjsCOlineList}; Na and K from \citet{BurrowsEtal2000apjBDspectra}; Rayleigh opacities from H, He, and {\molhyd} \citep{Kurucz_1970, LecavelierDesEtangsEtal2008aaRayleighHD189}; and collision-induced absorption from {\molhyd}--{\molhyd} \citep{BorysowEtal2001jqsrtH2H2highT, Borysow2002jqsrtH2H2lowT} and {\molhyd}--He \citep{BorysowEtal1988apjH2HeRT, borysowfrommhold1989b, borysowfrommhold1989a}.  Since the HITEMP and ExoMol databases consist of billions of line transitions, we first apply the repack package \citep{Cubillos2017apjCompress} to extract only the strong line transitions that dominate the opacity spectrum between 300 and 3000 K.  The final line list contains 63 million transitions.

We consider three metallicities for each planet in the sample: [M/H] = $-1$, 0, and 1; and two different Bond albedos: $A_{\rm B}$ = 0.0 and 0.3.  Overall, our radiative equilibrium temperature profiles are similar to those of \citet{MalikEtal2017ajHELIOS} and follow similar trends; for example, both display similar behaviours when we vary the atmospheric metallicity.

The Reproducible Research Compendium for the Pyrat Bay radiative equilibrium models of this article is available at \href{http://github.com/pcubillos/PassEtal2018\_EmissionSpectra}{github.com/pcubillos/PassEtal2018\_EmissionSpectra}.

\subsubsection{Simulated observations}

To simulate broad-band observations, each theoretical emission spectrum is converted from spectral radiance to a simulated eclipse depth using Equation~\ref{eq:eclipse}, neglecting the reflected-light term and assuming $R_\textrm{p}$/$R_*=0.1$ and \hbox{$T_{\textrm{b},*}=4000$ K}. A more appropriate treatment would use Kurucz models or vary $R_\textrm{p}$/$R_*$ and $T_{\textrm{b},*}$, although consideration of such complexities is time intensive. As we find that the simplified and intensive methods yield comparable results for a test subset of our spectra, we elect to use the simplified version for efficiency.  Eclipse depth is averaged over the band and Equations~\ref{eq:eclipse} and~\ref{eq:bright} are used to convert the band-integrated eclipse depth back to brightness temperature.

For benchmarking the performance of our GP regression, we consider the conservative case of 3.6 and \SI{4.5}{\micro\metre} broad-band measurements, corresponding to IRAC channels 1 and 2, as well as WFC3 spectral data. Following the precedent of  \citet{Stevenson_2017} and \citet{Keating_2017}, we elect not to include any simulated ground-based observations. As \citet{Croll_2015} discuss, uncertainties claimed for these measurements have historically been unreliable due to the systematics involved with near-IR, ground-based photometry and a lack of demonstrated repeatability. While \citet{Croll_2015} and \citet{Zhou_2015} show that uncertainties for ground-based eclipse measurements can be robustly estimated through careful systematic analysis and multiple independent measurements, there is not yet a critical mass of such repeated measurements in the literature. While we neglect ground-based observations in this analysis, our GP regression method can trivially be extended to an arbitrary number of observations in any ensemble of bands.

We consider a variety of binning options for the simulated WFC3 observations, although we find that the resolution of the spectral data does not significantly impact effective temperature estimation for observations taken over one eclipse.  For this reason, we elect to treat the WFC3 data as another broad-band measurement.

\subsubsection{Simulated uncertainties}

We can directly calculate uncertainties in our simulated measurements in the photon-limited regime, a good approximation for space-based observations.  As the flux is dominated by the star, we can approximate the number of photons in a given band using only the stellar spectral radiance:
\begin{equation}
N_{\rm phot} = \frac{\pi^2\tau\Delta t}{hc} \Bigg( \frac{R_*D}{2d}\Bigg)^2 \int_{\lambda_1}^{\lambda_2} B_\lambda(T_{\textrm{b},*})\lambda d\lambda,
\end{equation}
wherein $\tau$ is the system throughput, $D$ is the telescope diameter, $d$ is the distance to the star, and $\Delta t$ is the integration time.  The photon-limited precision is $\sqrt[]{2}$/$\sqrt[]{N_{\rm phot}}$, with the $\sqrt[]{2}$ factor resulting from the differential measurement of eclipse depth, assuming one acquires as much out-of-eclipse baseline as in-eclipse observations \citep{Cowan_2015}. We then propagate uncertainty in eclipse depth to uncertainty in spectral radiance. This uncertainty is simulated using a random Gaussian distribution before conversion to brightness temperature, as the uncertainty in brightness temperature is non-symmetric given symmetric uncertainties in flux.

We consider three physical scenarios to set uncertainty levels.  In the first, we assume photon-limited precision and a distance of $d=20$ pc, corresponding to an \hbox{HD 189733 b} analogue.  In the second, we place the planet at a distance of $d=245$ pc, the median value of the 43 planets in the exoplanet.org secondary eclipse sample with known distances. For the final case, we place the planet at $d=245$ pc and inflate the uncertainties in eclipse depth by a factor of two to represent pessimistic deviations from photon-limited precision \citep{Hansen_2014}. We assume an integration time of \hbox{$\Delta t=2$ hr}, corresponding to the typical duration of one eclipse \citep[Exoplanets Data Explorer,][]{Han_2014}. $D$ and $\tau$ are set to the published values for \textit{HST} WFC3 or $Spitzer$ IRAC, depending on the observation being simulated.

These three scenarios can be interpreted as best, good, and typical uncertainty levels; however, the signal-to-noise ratio (SNR) of the simulated eclipses can still vary significantly within a scenario as a result of hotter planets emitting more photons.  For this reason, we present our results grouped by SNR of the simulated \SI{4.5}{\micro\metre} eclipse depth rather than by physical scenario.  We consider eclipses with $\textrm{SNR} > 20$ as low uncertainty, with $5 < \textrm{SNR} < 20$ as medium uncertainty and with $2.5 < \textrm{SNR} < 5$ as high uncertainty.  Simulated eclipses with $\textrm{SNR} < 2.5$ are rejected from our analysis.  These categories contain 32\%, 23\%, 26\%, and 19\% of the simulated observations, respectively.

\subsection{Evaluation metric and method comparison}
\label{sec:methods-eval}
We consider 100 sets of simulated brightness temperatures for each of 324 models and three physical scenarios, resulting in 97,200 data sets.  First, we estimate effective temperature using the Error-Weighted Mean and Linear Interpolation methods.  For the EWM method, we average the brightness temperatures, weighting each observation by the inverse square of its uncertainty \citep{Schwartz_2015}. For the LI method, we set a constant brightness temperature shortward of the shortest wavelength measurement and longward of the longest wavelength measurement, linearly interpolating between the other measurements using SciPy's interp1d routine \citep{Cowan_2011}. The resulting $T_\textrm{b,p}(\lambda)$ spectrum is converted to an effective temperature using Equations~\ref{eq:teff} and \ref{eq:bright}.  For both methods, we calculate uncertainty using a 100-iteration Monte Carlo.

We then investigate the 97,200 data sets using GP regression. As before, we estimate uncertainties using a 100-iteration Monte Carlo (i.e., by performing GP regression on 100 instances of the data set).  We then sample 100 output functions from each fit; while the Monte Carlo accounts for the uncertainties in the observations, the latter approach accounts for the uncertainty in the GP result. The 10,000 brightness temperature functions are then converted to effective temperatures using Equations~\ref{eq:teff} and \ref{eq:bright}. Uncertainty is calculated as the standard deviation of the 10,000 effective temperature estimates.

\begin{figure}
	\includegraphics[width=1.2\columnwidth, center]{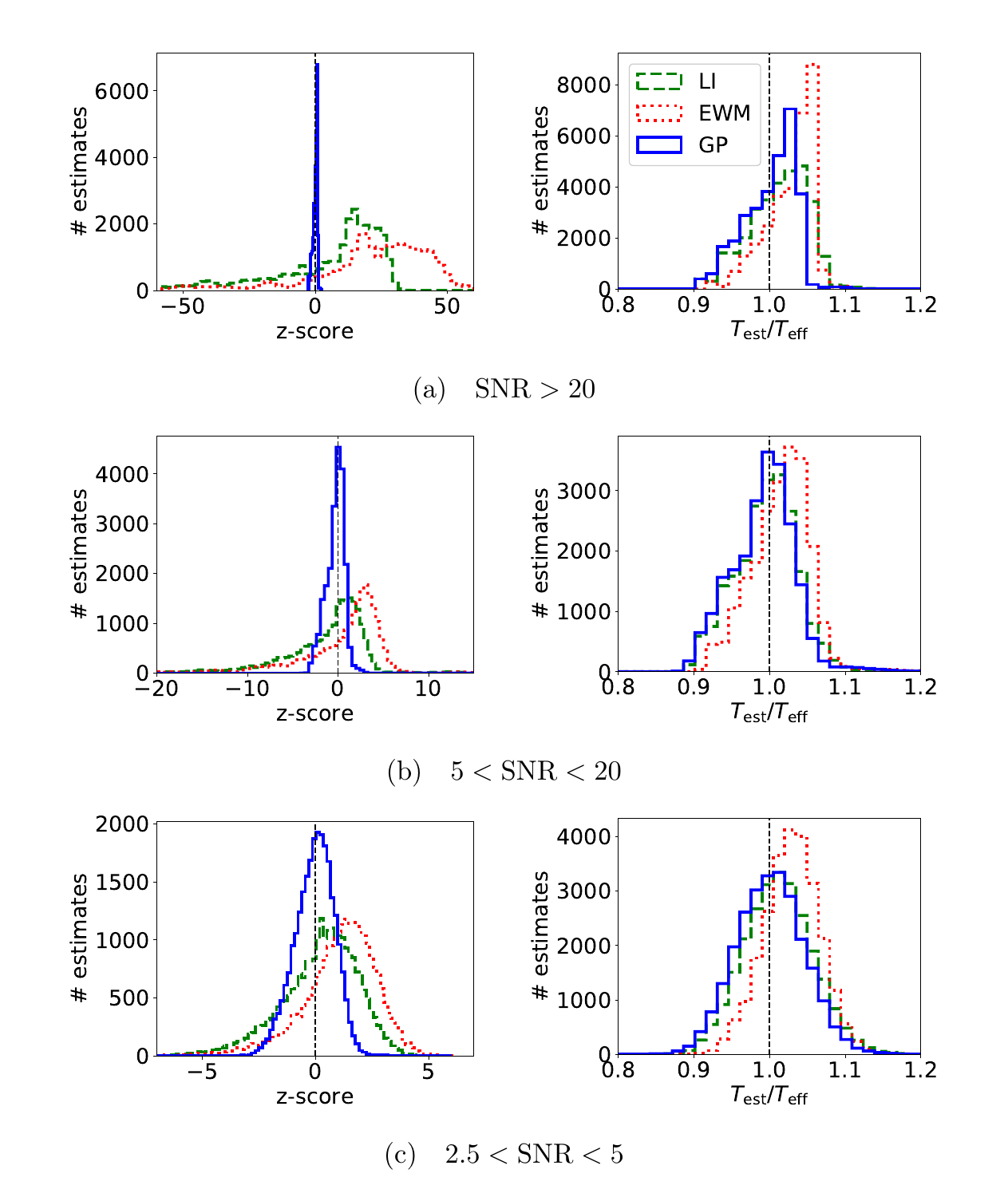}
    \caption{Distribution of z-scores and $T_{\rm est}/T_{\rm eff}$ for each of the EWM, GP, and LI methods, tested on the 97,200 data sets described in Section~\ref{sec:methods-eval} (with \emph{HST}/WFC3/G141, \emph{Spitzer}/IRAC/ch1, and \emph{Spitzer}/IRAC/ch2 observations). Results are grouped by the SNR of the \SI{4.5}{\micro\metre} eclipse depth. The z-score scale varies between panels to appropriately display the spread of the data.}
    \label{fig:zscores}
 \end{figure}

\begin{figure}
	\includegraphics[width=0.95\columnwidth]{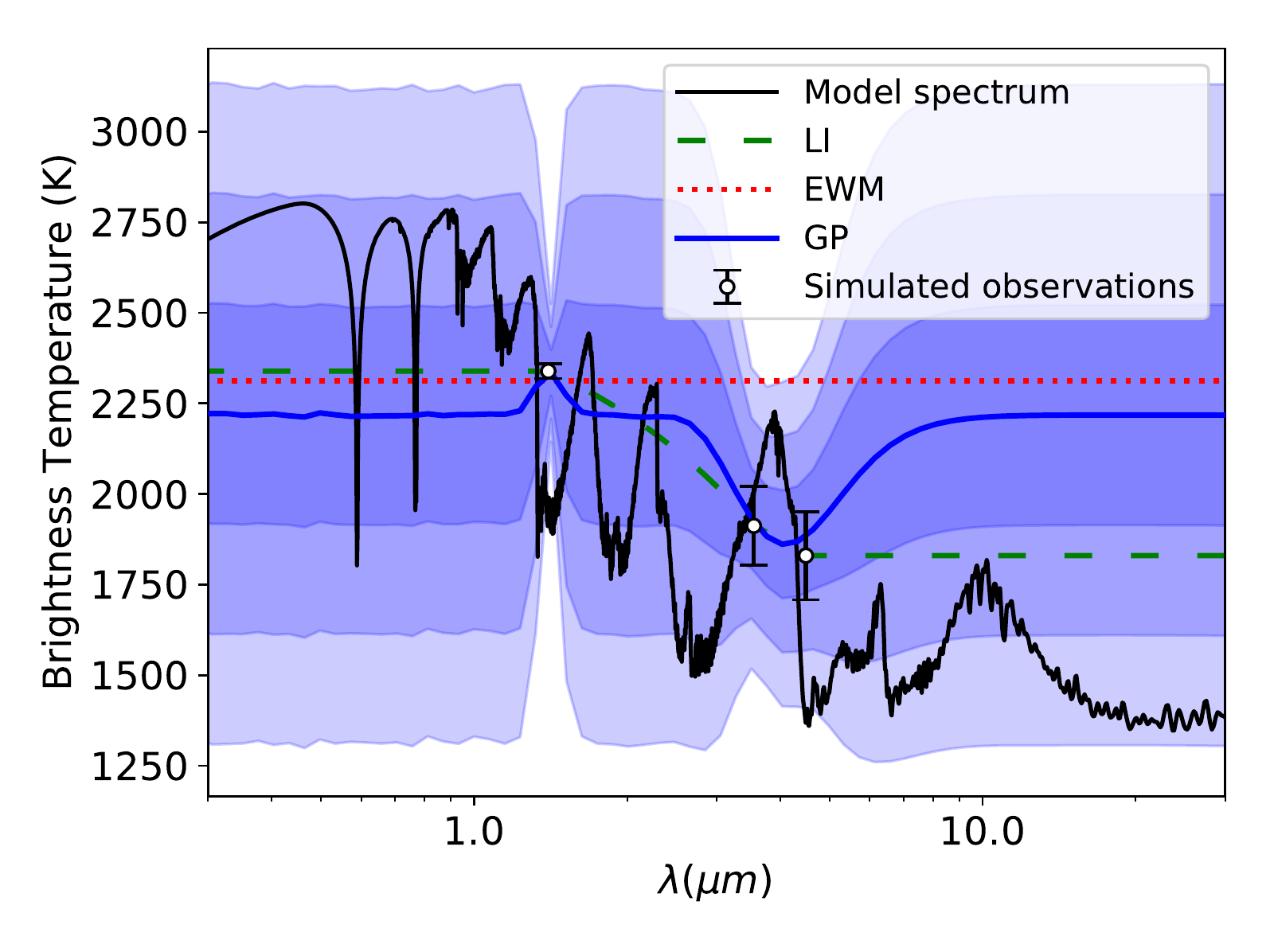}
    \caption{A model spectrum, simulated observations, and $T_\textrm{b,p}(\lambda)$ fits from the three methods. The contours show the 1$\sigma$, 2$\sigma$, and 3$\sigma$ confidence intervals for the GP fit.}
    \label{fig:fit}
\end{figure}

The three methods produce three effective temperature estimates ($T_{\rm est}$) for each of the 97,200 mock spectra, each with a corresponding uncertainty. We consider two metrics to benchmark the performance of these three estimation methods. The first is the ratio of the estimated effective temperature and the actual effective temperature ($T_{\rm est}/T_{\rm eff}$), shown in Figure 3 of \citet{Cowan_2011}; this metric allows us to directly compare the accuracy of the different methods. The second metric is the z-score:
\begin{equation}
\label{eq:z-score}
z = \frac{T_{\rm est}-T_{\rm eff}}{\sigma}.
\end{equation}
This metric evaluates the accuracy of the reported uncertainty, as well as the accuracy of the effective temperature estimate. A method that accurately estimates both the effective temperature and its uncertainty has a z-score distribution centered at zero with a standard deviation of one.
\section{Results}
\label{sec:results}
\subsection{Analysis of simulated data}
\label{sec:res-sim}

Our results are presented in Figure~\ref{fig:zscores}, with a sample fit shown in Figure~\ref{fig:fit}. Using the GP method, we find z-score distributions (mean $\pm$ standard deviation) of \hbox{$-0.05$ $\pm$ 0.85} ($\rm{SNR} > 20$), \hbox{$-0.29$ $\pm$ 0.99} ($5 < \rm{SNR} < 20$), and \hbox{$-0.04$ $\pm$ 0.87} ($2.5 < \rm{SNR} < 5$), all close to the desired distribution of \hbox{0 $\pm$ 1.}\footnote{The intermediate SNR regime has a mean z-score that is somewhat further from 0 than the other two. This is an artifact of our SNR regime definitions: we consider three physical scenarios (based on planet distance and deviations from photon-limited precision), but we report our results grouped by SNR. This means that each grouping is a superposition of discrete distributions, not a single, smooth distribution. The effect of this is particularly noticeable in the intermediate SNR regime, as it contains many planets from each of the three scenarios. If we instead perform our z-score analysis on the original scenario-grouped data (as opposed to the SNR-grouped data), we find that the mid-data quality scenario ($d$ = 245 pc) yields a z-score distribution that is more similar to the low SNR and high SNR regimes: -0.09 $\pm$ 0.89.}  The EWM and LI methods result in z-score distributions that are biased and that have standard deviations larger than 1.  This effect is most pronounced for simulated observations with high SNR, suggesting that these methods are capable of quantifying the error due to the uncertainty in individual data points, but not the error due to the spectrum's undersampling. The GP method considers both sources of uncertainty and hence performs accurately in all three SNR regimes.
 
While the GP method outperforms the LI and EWM methods in uncertainty estimation, all three methods produce reasonable effective temperature estimates, as evidenced by the similar $T_{\rm est}/T_{\rm eff}$ distributions in Figure~\ref{fig:zscores}.  The GP and LI methods in particular produce almost identical $T_{\rm est}/T_{\rm eff}$ distributions centred around 1, while the EWM is slightly biased in the positive direction. However, we find that switching the EWM method from $1/\sigma_i^2$ to $1/\sigma_i$ weighting (as motivated in Section~\ref{sec:gp-imp} and implemented in our GP method) reduces this bias. If the EWM is used in future work, we therefore recommend that it be used with $1/\sigma_i$ weighting.  However, we stress that the LI and EWM methods significantly underestimate the uncertainty in the effective temperature and hence should not be the preferred estimator for future work.

\begin{figure}
    \vspace{-0.3cm}
	\includegraphics[width=1.2\columnwidth, center]{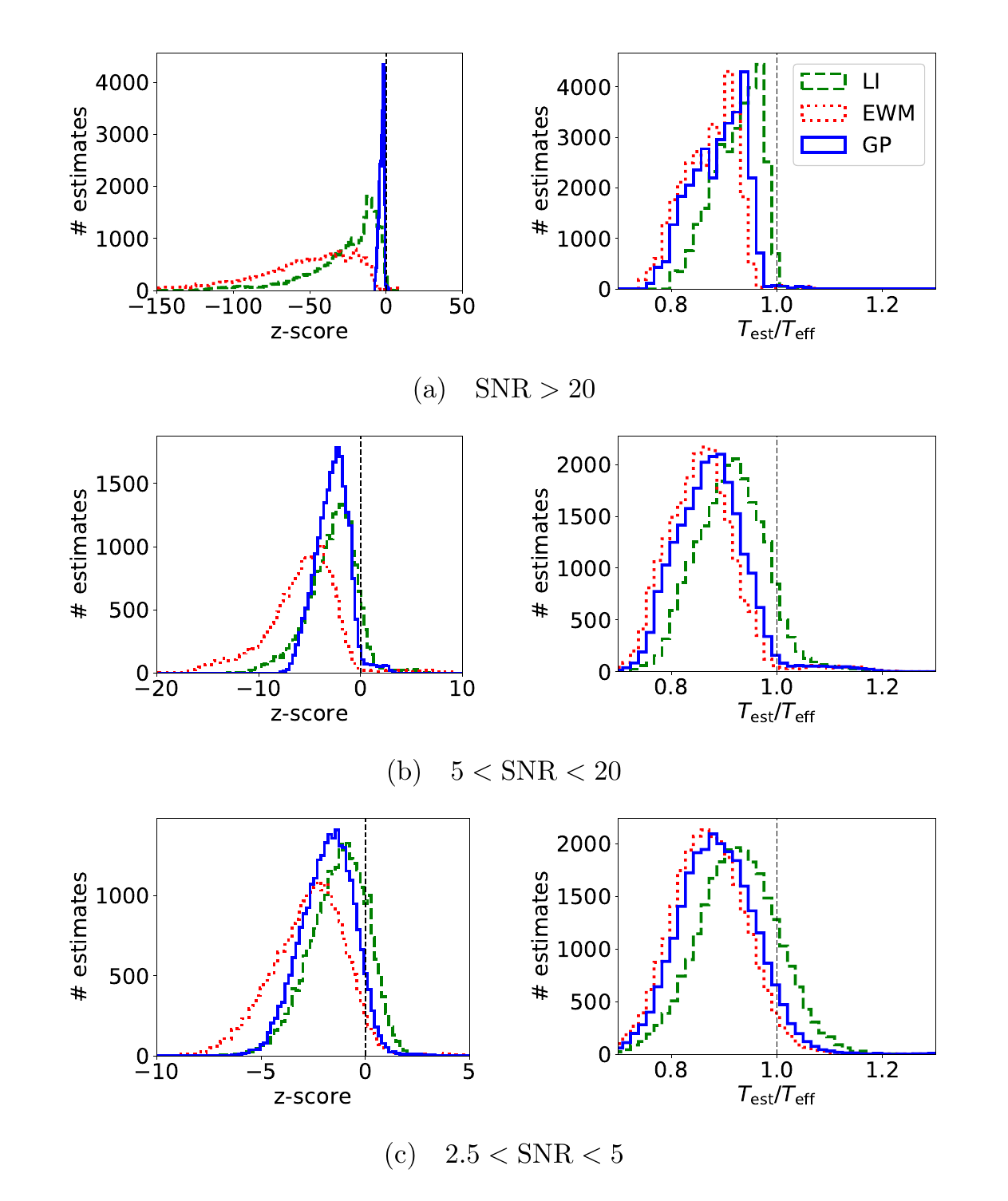}
	\vspace{-0.6cm}
    \caption{Distribution of z-scores and $T_{\rm est}/T_{\rm eff}$ for each of the EWM, GP, and LI methods, tested on the 97,200 data sets with observations only at 3.6 and \SI{4.5}{\micro\metre} (that is, without simulated WFC3 data). Results are grouped by the SNR of the \SI{4.5}{\micro\metre} eclipse depth and the z-score scale varies between panels to appropriately display the spread of the data.  In contrast to Figure~\ref{fig:zscores}, these results are skewed towards lower temperatures.}
    \label{fig:spitzer}
 \end{figure}

We have shown that the GP, LI, and EWM methods are capable of estimating the effective temperature of a planet with three observed eclipse depths (WFC3, IRAC channel 1, and IRAC channel 2) in an unbiased way, with the GP method able to accurately evaluate the uncertainties in these estimates.  As many planets have only been observed with warm \textit{Spitzer}, it may be tempting to try to extend these methods to data sets with only 3.6 and \SI{4.5}{\micro\metre} eclipse measurements.  However, such analyses should be undertaken with caution.  In Figure~\ref{fig:spitzer}, we show the z-score and $T_{\rm est}/T_{\rm eff}$ distributions for an analysis of the same 324 model spectra, but for which observations have only been simulated in the warm \textit{Spitzer} bands.  Both the z-score and $T_{\rm est}/T_{\rm eff}$ distributions are skewed left for all three estimation methods, indicating that the effective temperature is consistently underestimated when only the two IRAC bands are used.  This result is unsurprising: the 3.6 and \SI{4.5}{\micro\metre} bands will always emanate from higher in the atmosphere and hence represent the cooler part of the spectrum, as we do not include any models with thermal inversions in our analysis (see Figure~\ref{fig:fit}). An estimate made with only these two bands will therefore underestimate the planet's effective temperature. To a lesser extent, the same effect can also be seen in Figure~\ref{fig:zscores}: since we do not consider atmospheres with thermal inversions, our most certain data point (the WFC3 measurement) always represents a deeper, warmer layer in the atmosphere. Where the WFC3 measurement has a very small uncertainty (SNR > 20), we therefore see the $T_{\rm est}/T_{\rm eff}$ distribution slightly skewed towards overestimates of temperature, regardless of estimation method.

The results of Figure~\ref{fig:spitzer} can be seen as a worst-case scenario: in addition to only considering non-inverted atmospheres, the models evaluated here do not contain clouds, which would flatten the spectrum and hence reduce the amplitude of the bias.  However, an estimate of uncertainty should not assume significantly greater performance than this worst-case. We therefore conclude that the effective temperature is not known to better than 20\% at the 1$\sigma$ level if only 3.6 and \SI{4.5}{\micro\metre} eclipse depths have been observed.  This is in contrast to the results of \citet{Cowan_2011}, who claim a 1$\sigma$ systematic error of 5\% for observations taken in two bands (their Figure 3). However, their numerical experiments considered random combinations of broad-bands. The discrepancy in these results suggests that systematic error is strongly influenced by the particular wavelengths observed and not merely the number of observations.

\subsection{Analysis of archival data}
\label{sec:res-dat}

Having verified the accuracy of our GP estimates, we apply our method to archival data. The last systematic study of effective temperatures for archival hot Jupiters was performed by \citet{Cowan_2011}, which predates WFC3 observations. The uniform analysis we present here therefore provides a unique catalogue of effective temperatures, with no counterparts currently available elsewhere in the literature. Our catalogue can be interpreted as a hypothesis: motivated by our analysis of simulated data in Section~\ref{sec:res-sim}, we assert that 68\% of the time, the true effective temperatures of planets in our catalogue will fall within the 1$\sigma$ intervals we provide. This hypothesis will soon be testable, as upcoming missions such as the James Webb Space Telescope \citep[JWST;][]{Beichman_2014} and the Atmospheric Remote-sensing Infrared Exoplanet Large-survey \citep[ARIEL;][]{Puig_2016} will allow full secondary eclipse spectra to be measured for these hot Jupiters, unambiguously determining bolometric fluxes and dayside effective temperatures. However, our method will not be redundant in the age of JWST: since full spectral coverage will often require observations with multiple instruments and modes, there will always be targets that can benefit from a robust method for predicting the full bolometric flux from sparse observations.

A review of the literature yields thirteen hot Jupiters with WFC3 emission spectra, twelve of which have also been observed in the 3.6 and \SI{4.5}{\micro\metre} IRAC channels.  We neglect the thirteenth, \hbox{WASP-121 b} \citep{Evans_2017}, as it has not been observed at \SI{4.5}{\micro\metre}. The secondary eclipse depths for the twelve planets are listed in Table~\ref{tab:archive}.  Where available, we use the band-averaged WFC3 eclipse depth as published in the referenced paper.  Where only the full spectral data have been published, we take the average over the spectrum, determining uncertainties using a 1000-iteration Monte Carlo.  In the few cases where asymmetric uncertainties in the eclipse depth are given in the source paper, we list the largest of the two.  For \hbox{WASP-12 b}, we adopt the WFC3 eclipse depth from the \citet{Stevenson_2014b} reanalysis as opposed to that originally published in \citet{Swain_2013}.

\begin{table*}
\centering
\begin{tabular}{lllll}
\toprule
\multicolumn{1}{c}{\multirow{2}{*}{\textbf{Planet}}} & \multicolumn{3}{c}{ \textbf{$F_\textrm{p,day}/F_*$ (\%)}} & \multicolumn{1}{c}{\multirow{2}{*}{\textbf{T$_{\rm eff}$ (K)}}} \\
\cmidrule(lr){2-4}
 & \multicolumn{1}{c}{\textit{WFC3}} & \multicolumn{1}{c}{\textit{IRAC CH1}} & \multicolumn{1}{c}{\textit{IRAC CH2}} \\
 \midrule
CoRoT-2 b & 0.040\phantom{0} $\pm$ 0.007 $^{1}$ & 0.355 $\pm$ 0.020 $^{2}$ & 0.500 $\pm$ 0.020 $^{2}$ & 1846 $\pm$ \phantom{0}76 \\
HAT-P-7 b & 0.052\phantom{0} $\pm$ 0.001 $^{3}$ & 0.156 $\pm$ 0.009 $^{4}$ & 0.190 $\pm$ 0.006 $^{4}$ & 2775 $\pm$ \phantom{0}96 \\
HD 189733 b & 0.0096 $\pm$ 0.0039 $^{5}$ & 0.256 $\pm$ 0.014 $^{6}$ & 0.214 $\pm$ 0.020 $^{6}$ & 1490 $\pm$ \phantom{0}68  \\
HD 209458 b & 0.0082 $\pm$ 0.0015 $^{7}$ & 0.119 $\pm$ 0.007 $^{8}$ & 0.123 $\pm$ 0.006 $^{8}$ & 1523 $\pm$ \phantom{0}66\\
Kepler-13A b & 0.0734 $\pm$ 0.0028 $^{9}$ & 0.156 $\pm$ 0.031 $^{10}$ & 0.222 $\pm$ 0.023 $^{10}$ & 2918 $\pm$ 109 \\
TrES-3 b & 0.048\phantom{0} $\pm$ 0.007 $^{11}$& 0.356 $\pm$ 0.035 $^{12}$ & 0.372 $\pm$ 0.054 $^{12}$ & 1830 $\pm$ \phantom{0}83 \\
WASP-4 b & 0.063\phantom{0} $\pm$ 0.005 $^{11}$ & 0.319 $\pm$ 0.031 $^{13}$ & 0.343 $\pm$ 0.027 $^{13}$ & 1900 $\pm$ \phantom{0}79\\
WASP-12 b & 0.159\phantom{0} $\pm$ 0.004 $^{14,15}$ & 0.421 $\pm$ 0.011 $^{15}$ & 0.428 $\pm$ 0.012 $^{15}$ & 2899 $\pm$ 100 \\
WASP-18 b & 0.1045 $\pm$ 0.0006 $^{16}$ & 0.30\phantom{0} $\pm$ 0.02 $^{17}$ & 0.39\phantom{0} $\pm$ 0.02 $^{17}$ & 3067 $\pm$ 104 \\
WASP-33 b & 0.119\phantom{0} $\pm$ 0.006 $^{18}$ & 0.26\phantom{0} $\pm$ 0.05 $^{19}$ & 0.41\phantom{0} $\pm$ 0.02 $^{19}$ & 3108 $\pm$ 113 \\
WASP-43 b & 0.0456 $\pm$ 0.0010 $^{20}$ & 0.19\phantom{0} $\pm$ 0.01 $^{21}$ & 0.224 $\pm$ 0.018 $^{21}$ & 1464 $\pm$ \phantom{0}67 \\
WASP-103 b & 0.151\phantom{0} $\pm$ 0.015 $^{22}$ & 0.446 $\pm$ 0.38 $^{22}$ & 0.569 $\pm$ 0.014 $^{22}$ & 3205 $\pm$ 136 \\
\bottomrule
\end{tabular}
\caption{Published WFC3, IRAC channel 1, and IRAC channel 2 secondary eclipse depths alongside our GP-determined, model-independent dayside effective temperatures and associated uncertainties for the twelve planets in our archival analysis. \newline \textbf{References:} $^{1}$\citealt{Wilkins_2014}, $^{2}$\citealt{Deming_2011}, $^{3}$\citealt{Mansfield_2018}, $^{4}$\citealt{Wong_2016}, $^{5}$\citealt{Crouzet_2014}, $^{6}$\citealt{Charbonneau_2008}, $^{7}$\citealt{Line_2016}, $^{8}$\citealt{DiamondLowe_2014},
$^{9}$\citealt{Beatty_2017}, $^{10}$\citealt{Shporer_2014}, $^{11}$\citealt{Ranjan_2014}, $^{12}$\citealt{Fressin_2010}, $^{13}$\citealt{Beerer_2011}, $^{14}$\citealt{Swain_2013}, $^{15}$\citealt{Stevenson_2014b}, $^{16}$\citealt{Arcangeli_2018}, $^{17}$\citealt{Nymeyer_2011}, $^{18}$\citealt{Haynes_2015}, $^{19}$\citealt{Deming_2012}, $^{20}$\citealt{Stevenson_2014}, $^{21}$\citealt{Blecic_2013}, $^{22}$\citealt{Kreidberg_2018}}
\label{tab:archive}
\end{table*}

We apply our GP regression method to the twelve planets. As in Section~\ref{sec:gp-imp}, we use Kurucz models for stellar spectral radiance and adopt the system parameters (stellar $T_{\rm eff}$, stellar log($g$), and $R_\textrm{p}^2/R_*^2$) tabulated in the Exoplanet Data Explorer \citep[][accessed July 2018]{Han_2014}, with a stellar log($g$) of \hbox{4.71 $\pm$ 0.20 cms$^{-2}$} for CoRoT-2 \citep{Ammler-vonEiff_2009}.  Uncertainties in $T_{\rm b}$ are propagated from uncertainties in eclipse depth and system parameters using a 1000-iteration Monte Carlo.  The results of our analysis are listed in the final column of Table~\ref{tab:archive}.

\begin{figure}
    \vspace{-0.1cm}
	\includegraphics[width=\columnwidth]{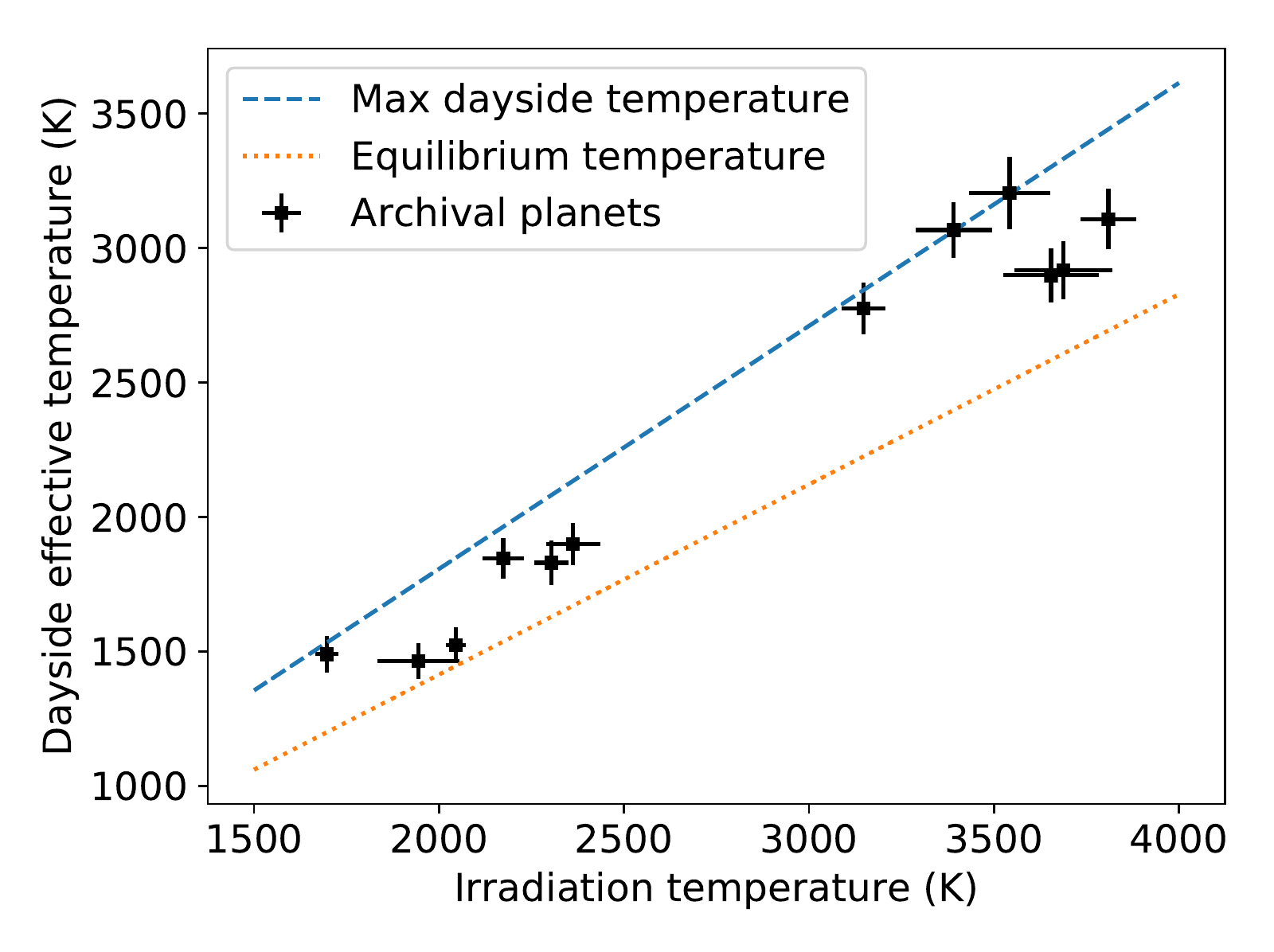}
	\vspace{-0.2cm}
    \caption{The irradiation temperatures and GP-estimated dayside effective temperatures for the twelve archival planets. In each case, we find effective temperatures consistent with imperfect day-night heat circulation.}
    \label{fig:irradiation}
\end{figure}

In Figure~\ref{fig:irradiation}, we show these GP-determined effective temperature estimates alongside each planet's irradiance temperature, $T_0 \equiv T_* \sqrt[]{R_*/a}$. Dashed lines are used to indicate equilibrium temperatures, $T_{\rm eq}=(1/4)^{1/4}T_0$, and maximal dayside temperatures, $T_{\rm max}=(2/3)^{1/4}T_0$; the former corresponds to a planet with zero Bond albedo and perfect heat circulation, while the latter corresponds to a planet with zero Bond albedo and negligible circulation. For the twelve archival planets, we find that our estimated dayside effective temperatures fall between these extremes, suggesting imperfect heat transport between the planets' day and night sides, and/or non-zero Bond albedo. Our results are consistent with \citet{Schwartz_2015}, who found that hotter planets generally deviate more strongly from equilibrium expectations (their Figure 2).

\section{Summary}
\label{sec:summary}
We have developed a Gaussian process method for estimating the dayside effective temperatures of hot Jupiters and their associated uncertainties, based on sparse secondary eclipse broad-band spectra. Our GP is trained on the HITEMP and exoplanets.org databases, from which we find a log-length scale of $-8.55$ (\hbox{$1.4\cdot10^{13}$ Hz}) and a log-signal variance of \hbox{$-4$} (\hbox{14\%}). With these hyperparameters, we are able to retrieve accurate effective temperature estimates for a diverse sample of simulated planets. The code for our method is publicly available on GitHub\footnote{\href{http://github.com/ekpass/gp-teff}{http://github.com/ekpass/gp-teff}} and can estimate a planet's effective temperature in seconds.

Through analysis of 97,200 sets of simulated WFC3 and warm \textit{Spitzer} observations, we show that the GP method is capable of producing unbiased effective temperature estimates with more accurate uncertainties than other model-independent methods (Figure~\ref{fig:zscores}, based on \textit{HST}/WFC3/G141, \textit{Spitzer}/IRAC/ch1, and \textit{Spitzer}/IRAC/ch2). However, we find significant bias with all methods when only IRAC channel 1 and channel 2 observations are used. We caution that effective temperature can only be constrained with 1$\sigma$ confidence to within 20\% of the true value in such instances (Figure~\ref{fig:spitzer}).

The other methods considered in our analysis---the Linear Interpolation method of \citet{Cowan_2011} and the Error-Weighted Mean method of \citet{Schwartz_2015}---estimate effective temperatures with similar accuracy, although they produce biased z-score distributions.  As error from the undersampling of the spectrum dominates in the high-SNR regime and such error is neglected by the EWM and LI methods, the poorest uncertainty estimates are produced for high-SNR data.  Additional uncertainty may be added in quadrature to account for this effect, as in \citet{Cowan_2011}, although these methods cannot dynamically determine undersampling's contribution. The GP method considers this uncertainty implicitly and hence appropriately estimates error in all SNR regimes. While we therefore recommend that the GP method replace the EWM method in general use, if the EWM is for some reason preferred, we advocate for the adoption of $1/\sigma_i$ error-weighting.  We find this produces more accurate, outlier-resistant results than the $1/\sigma_i^2$ weighting that has been used historically.

Having validated the GP method on simulated observations, we conclude with an analysis of archival data. In Table~\ref{tab:archive}, we provide effective temperature estimates with robustly determined uncertainties for the twelve hot Jupiters with published WFC3 and IRAC observations. These results may inform future statistical analyses of day-to-night heat transport and Bond albedo, in the vein of \citet{Cowan_2011} and \citet{Schwartz_2015}.
\section*{Acknowledgements}
E.K.P. is funded by an Institut de Recherche sur les Exoplan\`etes (iREx) Trottier Excellence Grant and a Natural Sciences and Engineering Research Council (NSERC) Undergraduate Student Research Award.  The authors thank Evelyn Macdonald and the anonymous referees for comments that improved this manuscript, and James Xu for preliminary comparisons of effective temperature estimators.



\bibliographystyle{mnras}
\bibliography{gp} 

\begin{thebibliography}{}
\makeatletter
\relax
\def\mn@urlcharsother{\let\do\@makeother \do\$\do\&\do\#\do\^\do\_\do\%\do\~}
\def\mn@doi{\begingroup\mn@urlcharsother \@ifnextchar [ {\mn@doi@}
  {\mn@doi@[]}}
\def\mn@doi@[#1]#2{\def\@tempa{#1}\ifx\@tempa\@empty \href
  {http://dx.doi.org/#2} {doi:#2}\else \href {http://dx.doi.org/#2} {#1}\fi
  \endgroup}
\def\mn@eprint#1#2{\mn@eprint@#1:#2::\@nil}
\def\mn@eprint@arXiv#1{\href {http://arxiv.org/abs/#1} {{\tt arXiv:#1}}}
\def\mn@eprint@dblp#1{\href {http://dblp.uni-trier.de/rec/bibtex/#1.xml}
  {dblp:#1}}
\def\mn@eprint@#1:#2:#3:#4\@nil{\def\@tempa {#1}\def\@tempb {#2}\def\@tempc
  {#3}\ifx \@tempc \@empty \let \@tempc \@tempb \let \@tempb \@tempa \fi \ifx
  \@tempb \@empty \def\@tempb {arXiv}\fi \@ifundefined
  {mn@eprint@\@tempb}{\@tempb:\@tempc}{\expandafter \expandafter \csname
  mn@eprint@\@tempb\endcsname \expandafter{\@tempc}}}

\bibitem[\protect\citeauthoryear{Almosallam, Lindsay, Jarvis  \&
  Roberts}{Almosallam et~al.}{2016}]{Almosallam_2016}
Almosallam I.~A.,  Lindsay S.~N.,  Jarvis M.~J.,   Roberts S.~J.,  2016,
  \mn@doi [\mnras] {10.1093/mnras/stv2425}, 455, 2387

\bibitem[\protect\citeauthoryear{Ambikasaran, Foreman-Mackey, Greengard, Hogg
  \& O'Neil}{Ambikasaran et~al.}{2015}]{Ambikasaran_2015}
Ambikasaran S.,  Foreman-Mackey D.,  Greengard L.,  Hogg D.~W.,   O'Neil M.,
  2015, \mn@doi [IEEE Transactions on Pattern Analysis and Machine
  Intelligence] {10.1109/TPAMI.2015.2448083}, 38

\bibitem[\protect\citeauthoryear{Ammler-von Eiff, Santos, Sousa, Fernandes,
  Guillot, Israelian, Mayor  \& Melo}{Ammler-von Eiff
  et~al.}{2009}]{Ammler-vonEiff_2009}
Ammler-von Eiff M.,  Santos N.~C.,  Sousa S.~G.,  Fernandes J.,  Guillot T.,
  Israelian G.,  Mayor M.,   Melo C.,  2009, \mn@doi [\aap]
  {10.1051/0004-6361/200912360}, 507, 523

\bibitem[\protect\citeauthoryear{Arcangeli et~al.,}{Arcangeli
  et~al.}{2018}]{Arcangeli_2018}
Arcangeli J.,  et~al., 2018, \mn@doi [\apjl] {10.3847/2041-8213/aab272}, 855,
  L30

\bibitem[\protect\citeauthoryear{{Asplund}, {Grevesse}, {Sauval}  \&
  {Scott}}{{Asplund} et~al.}{2009}]{AsplundEtal2009araSolarComposition}
{Asplund} M.,  {Grevesse} N.,  {Sauval} A.~J.,   {Scott} P.,  2009, \mn@doi
  [\araa] {10.1146/annurev.astro.46.060407.145222}, \href
  {http://adsabs.harvard.edu/abs/2009ARA\%26A..47..481A} {47, 481}

\bibitem[\protect\citeauthoryear{Aster, Borchers  \& Thurber}{Aster
  et~al.}{2013}]{Aster_2013}
Aster R.~C.,  Borchers B.,   Thurber C.~H.,  2013, Parameter Estimation and
  Inverse Problems.
Elsevier

\bibitem[\protect\citeauthoryear{Bahcall}{Bahcall}{1986}]{Bahcall_1986}
Bahcall N.~A.,  1986, \mn@doi [Annals of the New York Academy of Sciences]
  {10.1111/j.1749-6632.1986.tb47983.x}, 470, 331

\bibitem[\protect\citeauthoryear{Barman, Hauschildt  \& Allard}{Barman
  et~al.}{2005}]{Barman_2005}
Barman T.~S.,  Hauschildt P.~H.,   Allard F.,  2005, \mn@doi [\apj]
  {10.1086/444349}, 632, 1132

\bibitem[\protect\citeauthoryear{Beatty, Madhusudhan, Tsiaras, Zhao, Gilliland,
  Knutson, Shporer  \& Wright}{Beatty et~al.}{2017}]{Beatty_2017}
Beatty T.~G.,  Madhusudhan N.,  Tsiaras A.,  Zhao M.,  Gilliland R.~L.,
  Knutson H.~A.,  Shporer A.,   Wright J.~T.,  2017, \mn@doi [\aj]
  {10.3847/1538-3881/aa899b}, 154, 158

\bibitem[\protect\citeauthoryear{Beerer et~al.,}{Beerer
  et~al.}{2011}]{Beerer_2011}
Beerer I.~M.,  et~al., 2011, \mn@doi [\apj] {10.1088/0004-637X/727/1/23}, 727,
  23

\bibitem[\protect\citeauthoryear{Beichman et~al.,}{Beichman
  et~al.}{2014}]{Beichman_2014}
Beichman C.,  et~al., 2014, \mn@doi [\pasp] {10.1086/679566}, 126, 1134

\bibitem[\protect\citeauthoryear{Benneke}{Benneke}{2015}]{Benneke_2015}
Benneke B.,  2015, preprint (\mn@eprint {arXiv} {1504.07655})

\bibitem[\protect\citeauthoryear{{Blecic}}{{Blecic}}{2016}]{Blecic2016phdThesis}
{Blecic} J.,  2016, preprint (\mn@eprint {arXiv} {1604.02692})

\bibitem[\protect\citeauthoryear{Blecic et~al.,}{Blecic
  et~al.}{2013}]{Blecic_2013}
Blecic J.,  et~al., 2013, \mn@doi [\apj] {10.1088/0004-637X/779/1/5}, 779, 5

\bibitem[\protect\citeauthoryear{{Blecic}, {Harrington}  \& {Bowman}}{{Blecic}
  et~al.}{2016}]{BlecicEtal2016apsjTEA}
{Blecic} J.,  {Harrington} J.,   {Bowman} M.~O.,  2016, \mn@doi [\apjs]
  {10.3847/0067-0049/225/1/4}, \href
  {http://adsabs.harvard.edu/abs/2016ApJS..225....4B} {225, 4}

\bibitem[\protect\citeauthoryear{Blecic, Dobbs-Dixon  \& Greene}{Blecic
  et~al.}{2017}]{Blecic_2017}
Blecic J.,  Dobbs-Dixon I.,   Greene T.,  2017, \mn@doi [\apj]
  {10.3847/1538-4357/aa8171}, 848, 127

\bibitem[\protect\citeauthoryear{{Borysow}}{{Borysow}}{2002}]{Borysow2002jqsrtH2H2lowT}
{Borysow} A.,  2002, \mn@doi [\aap] {10.1051/0004-6361:20020555}, \href
  {http://adsabs.harvard.edu/abs/2002A\%26A...390..779B} {390, 779}

\bibitem[\protect\citeauthoryear{{Borysow} \& {Frommhold}}{{Borysow} \&
  {Frommhold}}{1989}]{borysowfrommhold1989a}
{Borysow} A.,  {Frommhold} L.,  1989, \mn@doi [\apj] {10.1086/167515}, \href
  {http://adsabs.harvard.edu/abs/1989ApJ...341..549B} {341, 549}

\bibitem[\protect\citeauthoryear{{Borysow}, {Frommhold}  \&
  {Birnbaum}}{{Borysow} et~al.}{1988}]{BorysowEtal1988apjH2HeRT}
{Borysow} J.,  {Frommhold} L.,   {Birnbaum} G.,  1988, \mn@doi [\apj]
  {10.1086/166112}, \href {http://adsabs.harvard.edu/abs/1988ApJ...326..509B}
  {326, 509}

\bibitem[\protect\citeauthoryear{{Borysow}, {Frommhold}  \&
  {Moraldi}}{{Borysow} et~al.}{1989}]{borysowfrommhold1989b}
{Borysow} A.,  {Frommhold} L.,   {Moraldi} M.,  1989, \mn@doi [\apj]
  {10.1086/167027}, \href {http://adsabs.harvard.edu/abs/1989ApJ...336..495B}
  {336, 495}

\bibitem[\protect\citeauthoryear{{Borysow}, {Jorgensen}  \& {Fu}}{{Borysow}
  et~al.}{2001}]{BorysowEtal2001jqsrtH2H2highT}
{Borysow} A.,  {Jorgensen} U.~G.,   {Fu} Y.,  2001, \mn@doi [\jqsrt]
  {10.1016/S0022-4073(00)00023-6}, \href
  {http://adsabs.harvard.edu/abs/2001JQSRT..68..235B} {68, 235}

\bibitem[\protect\citeauthoryear{Brewer \& Stello}{Brewer \&
  Stello}{2009}]{Brewer_2009}
Brewer B.~J.,  Stello D.,  2009, \mn@doi [\mnras]
  {10.1111/j.1365-2966.2009.14679.x}, 395, 2226

\bibitem[\protect\citeauthoryear{{Burrows}, {Marley}  \& {Sharp}}{{Burrows}
  et~al.}{2000}]{BurrowsEtal2000apjBDspectra}
{Burrows} A.,  {Marley} M.~S.,   {Sharp} C.~M.,  2000, \mn@doi [\apj]
  {10.1086/308462}, \href {http://adsabs.harvard.edu/abs/2000ApJ...531..438B}
  {531, 438}

\bibitem[\protect\citeauthoryear{{Cartier} et~al.,}{{Cartier}
  et~al.}{2017}]{Cartier_2017}
{Cartier} K.~M.~S.,  et~al., 2017, \mn@doi [\aj] {10.3847/1538-3881/153/1/34},
  \href {http://adsabs.harvard.edu/abs/2017AJ....153...34C} {153, 34}

\bibitem[\protect\citeauthoryear{{Castelli} \& {Kurucz}}{{Castelli} \&
  {Kurucz}}{2004}]{Castelli_2004}
{Castelli} F.,  {Kurucz} R.~L.,  2004, preprint (\mn@eprint {arXiv}
  {astro-ph/0405087})

\bibitem[\protect\citeauthoryear{Charbonneau, Knutson, Barman, Allen, Mayor,
  Megeath, Queloz  \& Udry}{Charbonneau et~al.}{2008}]{Charbonneau_2008}
Charbonneau D.,  Knutson H.~A.,  Barman T.,  Allen L.~E.,  Mayor M.,  Megeath
  S.~T.,  Queloz D.,   Udry S.,  2008, \mn@doi [\apj] {10.1086/591635}, 686,
  1341

\bibitem[\protect\citeauthoryear{Cheng et~al.,}{Cheng
  et~al.}{2000}]{Cheng_2000}
Cheng E.~S.,  et~al., 2000, in {Breckinridge} J.~B.,  {Jakobsen} P.,  eds,
  \procspie Vol. 4013, UV, Optical, and IR Space Telescopes and Instruments. pp
  367--373, \mn@doi{10.1117/12.394020}

\bibitem[\protect\citeauthoryear{Cowan \& Agol}{Cowan \&
  Agol}{2011}]{Cowan_2011}
Cowan N.~B.,  Agol E.,  2011, \mn@doi [\apj] {10.1088/0004-637X/729/1/54}, 729,
  54

\bibitem[\protect\citeauthoryear{Cowan, Agol  \& Charbonneau}{Cowan
  et~al.}{2007}]{Cowan_2007}
Cowan N.~B.,  Agol E.,   Charbonneau D.,  2007, \mn@doi [\mnras]
  {10.1111/j.1365-2966.2007.11897.x}, 379, 641

\bibitem[\protect\citeauthoryear{Cowan et~al.,}{Cowan
  et~al.}{2015}]{Cowan_2015}
Cowan N.~B.,  et~al., 2015, \mn@doi [\pasp] {10.1086/680855}, 127, 311

\bibitem[\protect\citeauthoryear{Croll et~al.,}{Croll
  et~al.}{2015}]{Croll_2015}
Croll B.,  et~al., 2015, \mn@doi [\apj] {10.1088/0004-637X/802/1/28}, 802, 28

\bibitem[\protect\citeauthoryear{Crossfield, Hansen  \& Barman}{Crossfield
  et~al.}{2012}]{Crossfield_2012}
Crossfield I. J.~M.,  Hansen B. M.~S.,   Barman T.,  2012, \mn@doi [\apj]
  {10.1088/0004-637X/746/1/46}, 746, 46

\bibitem[\protect\citeauthoryear{Crouzet, McCullough, Deming  \&
  Madhusudhan}{Crouzet et~al.}{2014}]{Crouzet_2014}
Crouzet N.,  McCullough P.~R.,  Deming D.,   Madhusudhan N.,  2014, \mn@doi
  [\apj] {10.1088/0004-637X/795/2/166}, 795, 166

\bibitem[\protect\citeauthoryear{{Cubillos}}{{Cubillos}}{2016}]{Cubillos2016phdThesis}
{Cubillos} P.~E.,  2016, preprint (\mn@eprint {arXiv} {1604.01320})

\bibitem[\protect\citeauthoryear{{Cubillos}}{{Cubillos}}{2017}]{Cubillos2017apjCompress}
{Cubillos} P.~E.,  2017, \mn@doi [\apj] {10.3847/1538-4357/aa9228}, \href
  {http://adsabs.harvard.edu/abs/2017ApJ...850...32C} {850, 32}

\bibitem[\protect\citeauthoryear{Deming, Harrington, Laughlin, Seager, Navarro,
  Bowman  \& Horning}{Deming et~al.}{2007}]{Deming_2007}
Deming D.,  Harrington J.,  Laughlin G.,  Seager S.,  Navarro S.~B.,  Bowman
  W.~C.,   Horning K.,  2007, \mn@doi [\apjl] {10.1086/522496}, 667, L199

\bibitem[\protect\citeauthoryear{Deming et~al.,}{Deming
  et~al.}{2011}]{Deming_2011}
Deming D.,  et~al., 2011, \mn@doi [\apj] {10.1088/0004-637X/726/2/95}, 726, 95

\bibitem[\protect\citeauthoryear{Deming et~al.,}{Deming
  et~al.}{2012}]{Deming_2012}
Deming D.,  et~al., 2012, \mn@doi [\apj] {10.1088/0004-637X/754/2/106}, 754,
  106

\bibitem[\protect\citeauthoryear{Diamond-Lowe, Stevenson, Bean, Line  \&
  Fortney}{Diamond-Lowe et~al.}{2014}]{DiamondLowe_2014}
Diamond-Lowe H.,  Stevenson K.~B.,  Bean J.~L.,  Line M.~R.,   Fortney J.~J.,
  2014, \mn@doi [\apj] {10.1088/0004-637X/796/1/66}, 796, 66

\bibitem[\protect\citeauthoryear{Evans, Aigrain, Gibson, Barstow, Amundsen,
  Tremblin  \& Mourier}{Evans et~al.}{2015}]{Evans_2015}
Evans T.~M.,  Aigrain S.,  Gibson N.,  Barstow J.~K.,  Amundsen D.~S.,
  Tremblin P.,   Mourier P.,  2015, \mn@doi [\mnras] {10.1093/mnras/stv910},
  451, 680

\bibitem[\protect\citeauthoryear{Evans et~al.,}{Evans
  et~al.}{2017}]{Evans_2017}
Evans T.~M.,  et~al., 2017, \mn@doi [\nat] {10.1038/nature23266}, 548, 58

\bibitem[\protect\citeauthoryear{Fazio et~al.,}{Fazio
  et~al.}{2004}]{Fazio_2004}
Fazio G.~G.,  et~al., 2004, \mn@doi [\apjs] {10.1086/422843}, 154, 10

\bibitem[\protect\citeauthoryear{Feng, Line, Fortney, Stevenson, Bean,
  Kreidberg  \& Parmentier}{Feng et~al.}{2016}]{Feng_2016}
Feng Y.~K.,  Line M.~R.,  Fortney J.~J.,  Stevenson K.~B.,  Bean J.,  Kreidberg
  L.,   Parmentier V.,  2016, \mn@doi [\apj] {10.3847/0004-637X/829/1/52}, 829,
  52

\bibitem[\protect\citeauthoryear{Foreman-Mackey, Hogg  \&
  Morton}{Foreman-Mackey et~al.}{2014}]{Foreman-Mackey_2014}
Foreman-Mackey D.,  Hogg D.~W.,   Morton T.~D.,  2014, \mn@doi [\apj]
  {10.1088/0004-637X/795/1/64}, 795, 64

\bibitem[\protect\citeauthoryear{Foreman-Mackey, Agol, Ambikasaran  \&
  Angus}{Foreman-Mackey et~al.}{2017}]{Foreman-Mackey_2017}
Foreman-Mackey D.,  Agol E.,  Ambikasaran S.,   Angus R.,  2017, \mn@doi [\aj]
  {10.3847/1538-3881/aa9332}, 154, 220

\bibitem[\protect\citeauthoryear{Fortney}{Fortney}{2018}]{Fortney_2018}
Fortney J.~J.,  2018, preprint (\mn@eprint {} {1804.08149})

\bibitem[\protect\citeauthoryear{Fortney, Marley, Lodders, Saumon  \&
  Freedman}{Fortney et~al.}{2005}]{Fortney_2005}
Fortney J.~J.,  Marley M.~S.,  Lodders K.,  Saumon D.,   Freedman R.,  2005,
  \mn@doi [\apjl] {10.1086/431952}, 627, L69

\bibitem[\protect\citeauthoryear{Fortney, Cooper, Showman, Marley  \&
  Freedman}{Fortney et~al.}{2006}]{Fortney_2006}
Fortney J.~J.,  Cooper C.~S.,  Showman A.~P.,  Marley M.~S.,   Freedman R.~S.,
  2006, \mn@doi [\apj] {10.1086/508442}, 652, 746

\bibitem[\protect\citeauthoryear{{Fortney}, {Marley}  \& {Barnes}}{{Fortney}
  et~al.}{2007}]{Fortney_2007}
{Fortney} J.~J.,  {Marley} M.~S.,   {Barnes} J.~W.,  2007, \mn@doi [\apj]
  {10.1086/512120}, \href
  {https://ui.adsabs.harvard.edu/abs/2007ApJ...659.1661F} {659, 1661}

\bibitem[\protect\citeauthoryear{Fressin, Knutson, Charbonneau, O'Donovan,
  Burrows, Deming, Mandushev  \& Spiegel}{Fressin et~al.}{2010}]{Fressin_2010}
Fressin F.,  Knutson H.~A.,  Charbonneau D.,  O'Donovan F.~T.,  Burrows A.,
  Deming D.,  Mandushev G.,   Spiegel D.,  2010, \mn@doi [\apj]
  {10.1088/0004-637X/711/1/374}, 711, 374

\bibitem[\protect\citeauthoryear{Gandhi \& Madhusudhan}{Gandhi \&
  Madhusudhan}{2018}]{Gandhi_2018}
Gandhi S.,  Madhusudhan N.,  2018, \mn@doi [\mnras] {10.1093/mnras/stx2748},
  474, 271

\bibitem[\protect\citeauthoryear{{Garhart} et~al.,}{{Garhart}
  et~al.}{2019}]{Garhart_2019}
{Garhart} E.,  et~al., 2019, preprint (\mn@eprint {arXiv} {1901.07040})

\bibitem[\protect\citeauthoryear{Gibson, Aigrain, Roberts, Evans, Osborne  \&
  Pont}{Gibson et~al.}{2012}]{Gibson_2012}
Gibson N.~P.,  Aigrain S.,  Roberts S.,  Evans T.~M.,  Osborne M.,   Pont F.,
  2012, \mn@doi [\mnras] {10.1111/j.1365-2966.2011.19915.x}, 419, 2683

\bibitem[\protect\citeauthoryear{Han, Wang, Wright, Feng, Zhao, Fakhouri, Brown
   \& Hancock}{Han et~al.}{2014}]{Han_2014}
Han E.,  Wang S.~X.,  Wright J.~T.,  Feng Y.~K.,  Zhao M.,  Fakhouri O.,  Brown
  J.~I.,   Hancock C.,  2014, \mn@doi [\pasp] {10.1086/678447}, 126, 827

\bibitem[\protect\citeauthoryear{Hansen, Schwartz  \& Cowan}{Hansen
  et~al.}{2014}]{Hansen_2014}
Hansen C.~J.,  Schwartz J.~C.,   Cowan N.~B.,  2014, \mn@doi [\mnras]
  {10.1093/mnras/stu1699}, 444, 3632

\bibitem[\protect\citeauthoryear{Haynes, Mandell, Madhusudhan, Deming  \&
  Knutson}{Haynes et~al.}{2015}]{Haynes_2015}
Haynes K.,  Mandell A.~M.,  Madhusudhan N.,  Deming D.,   Knutson H.,  2015,
  \mn@doi [\apj] {10.1088/0004-637X/806/2/146}, 806, 146

\bibitem[\protect\citeauthoryear{Henderson, Skemer, Morley  \&
  Fortney}{Henderson et~al.}{2017}]{Henderson_2017}
Henderson C.~S.,  Skemer A.~J.,  Morley C.~V.,   Fortney J.~J.,  2017, \mn@doi
  [\mnras] {10.1093/mnras/stx1495}, 470, 4557

\bibitem[\protect\citeauthoryear{Ingalls et~al.,}{Ingalls
  et~al.}{2016}]{Ingalls_2016}
Ingalls J.~G.,  et~al., 2016, \mn@doi [\aj] {10.3847/0004-6256/152/2/44}, 152,
  44

\bibitem[\protect\citeauthoryear{Irwin et~al.,}{Irwin
  et~al.}{2008}]{Irwin_2008}
Irwin P. G.~J.,  et~al., 2008, \mn@doi [\jqsrt] {10.1016/j.jqsrt.2007.11.006},
  109, 1136

\bibitem[\protect\citeauthoryear{Jones, Oliphant, Peterson  et~al.}{Jones
  et~al.}{2001}]{Jones_2001}
Jones E.,  Oliphant T.,  Peterson P.,   et~al., 2001, {SciPy}: Open source
  scientific tools for {Python}, \url {http://www.scipy.org/}

\bibitem[\protect\citeauthoryear{Kammer et~al.,}{Kammer
  et~al.}{2015}]{Kammer_2015}
Kammer J.~A.,  et~al., 2015, \mn@doi [\apj] {10.1088/0004-637X/810/2/118}, 810,
  118

\bibitem[\protect\citeauthoryear{Keating \& Cowan}{Keating \&
  Cowan}{2017}]{Keating_2017}
Keating D.,  Cowan N.~B.,  2017, \mn@doi [\apjl] {10.3847/2041-8213/aa8b6b},
  849, L5

\bibitem[\protect\citeauthoryear{Kreidberg et~al.,}{Kreidberg
  et~al.}{2018}]{Kreidberg_2018}
Kreidberg L.,  et~al., 2018, \mn@doi [\aj] {10.3847/1538-3881/aac3df}, 156, 17

\bibitem[\protect\citeauthoryear{{Kurucz}}{{Kurucz}}{1970}]{Kurucz_1970}
{Kurucz} R.~L.,  1970, SAO Special Report, \href
  {http://adsabs.harvard.edu/abs/1970SAOSR.309.....K} {309}

\bibitem[\protect\citeauthoryear{Lavie et~al.,}{Lavie
  et~al.}{2017}]{Lavie_2017}
Lavie B.,  et~al., 2017, \mn@doi [\aj] {10.3847/1538-3881/aa7ed8}, 154, 91

\bibitem[\protect\citeauthoryear{{Lecavelier Des Etangs}, {Pont},
  {Vidal-Madjar}  \& {Sing}}{{Lecavelier Des Etangs}
  et~al.}{2008}]{LecavelierDesEtangsEtal2008aaRayleighHD189}
{Lecavelier Des Etangs} A.,  {Pont} F.,  {Vidal-Madjar} A.,   {Sing} D.,  2008,
  \mn@doi [\aap] {10.1051/0004-6361:200809388}, \href
  {http://adsabs.harvard.edu/abs/2008A\%26A...481L..83L} {481, L83}

\bibitem[\protect\citeauthoryear{{Lee}, {Dobbs-Dixon}, {Helling}, {Bognar}  \&
  {Woitke}}{{Lee} et~al.}{2016}]{Lee_2016}
{Lee} G.,  {Dobbs-Dixon} I.,  {Helling} C.,  {Bognar} K.,   {Woitke} P.,  2016,
  \mn@doi [\aap] {10.1051/0004-6361/201628606}, \href
  {http://adsabs.harvard.edu/abs/2016A%26A...594A..48L} {594, A48}

\bibitem[\protect\citeauthoryear{{Lee}, {Wood}, {Dobbs-Dixon}, {Rice}  \&
  {Helling}}{{Lee} et~al.}{2017}]{Lee_2017}
{Lee} G.~K.~H.,  {Wood} K.,  {Dobbs-Dixon} I.,  {Rice} A.,   {Helling} C.,
  2017, \mn@doi [\aap] {10.1051/0004-6361/201629804}, \href
  {http://adsabs.harvard.edu/abs/2017A%26A...601A..22L} {601, A22}

\bibitem[\protect\citeauthoryear{{Li}, {Gordon}, {Rothman}, {Tan}, {Hu},
  {Kassi}, {Campargue}  \& {Medvedev}}{{Li}
  et~al.}{2015}]{LiEtal2015apjsCOlineList}
{Li} G.,  {Gordon} I.~E.,  {Rothman} L.~S.,  {Tan} Y.,  {Hu} S.-M.,  {Kassi}
  S.,  {Campargue} A.,   {Medvedev} E.~S.,  2015, \mn@doi [\apjs]
  {10.1088/0067-0049/216/1/15}, \href
  {http://adsabs.harvard.edu/abs/2015ApJS..216...15L} {216, 15}

\bibitem[\protect\citeauthoryear{Line et~al.,}{Line et~al.}{2013}]{Line_2013}
Line M.~R.,  et~al., 2013, \mn@doi [\apj] {10.1088/0004-637X/775/2/137}, 775,
  137

\bibitem[\protect\citeauthoryear{Line et~al.,}{Line et~al.}{2016}]{Line_2016}
Line M.~R.,  et~al., 2016, \mn@doi [\aj] {10.3847/0004-6256/152/6/203}, 152,
  203

\bibitem[\protect\citeauthoryear{Madhusudhan}{Madhusudhan}{2018}]{Madhusudhan_2018}
Madhusudhan N.,  2018, Atmospheric Retrieval of Exoplanets.
Springer International Publishing, Cham, pp 1--30

\bibitem[\protect\citeauthoryear{{Malik} et~al.,}{{Malik}
  et~al.}{2017}]{MalikEtal2017ajHELIOS}
{Malik} M.,  et~al., 2017, \mn@doi [\aj] {10.3847/1538-3881/153/2/56}, \href
  {http://adsabs.harvard.edu/abs/2017AJ....153...56M} {153, 56}

\bibitem[\protect\citeauthoryear{Mancini et~al.,}{Mancini
  et~al.}{2013}]{Mancini_2013}
Mancini L.,  et~al., 2013, \mn@doi [\mnras] {10.1093/mnras/stt1394}, 436, 2

\bibitem[\protect\citeauthoryear{Mansfield et~al.,}{Mansfield
  et~al.}{2018}]{Mansfield_2018}
Mansfield M.,  et~al., 2018, \mn@doi [\aj] {10.3847/1538-3881/aac497}, 156, 10

\bibitem[\protect\citeauthoryear{Marley, Gelino, Stephens, Lunine  \&
  Freedman}{Marley et~al.}{1999}]{Marley_1999}
Marley M.~S.,  Gelino C.,  Stephens D.,  Lunine J.~I.,   Freedman R.,  1999,
  \mn@doi [\apj] {10.1086/306881}, 513, 879

\bibitem[\protect\citeauthoryear{M{\'a}rquez-Neila, Fisher, Sznitman  \&
  Heng}{M{\'a}rquez-Neila et~al.}{2018}]{MarquezNeila_2018}
M{\'a}rquez-Neila P.,  Fisher C.,  Sznitman R.,   Heng K.,  2018, \mn@doi
  [Nature Astronomy] {10.1038/s41550-018-0504-2}

\bibitem[\protect\citeauthoryear{Martioli et~al.,}{Martioli
  et~al.}{2018}]{Martioli_2018}
Martioli E.,  et~al., 2018, \mn@doi [\mnras] {10.1093/mnras/stx3009}, 474, 4264

\bibitem[\protect\citeauthoryear{Morley, Knutson, Line, Fortney, Thorngren,
  Marley, Teal  \& Lupu}{Morley et~al.}{2017}]{Morley_2017}
Morley C.~V.,  Knutson H.,  Line M.,  Fortney J.~J.,  Thorngren D.,  Marley
  M.~S.,  Teal D.,   Lupu R.,  2017, \mn@doi [\aj]
  {10.3847/1538-3881/153/2/86}, 153, 86

\bibitem[\protect\citeauthoryear{Nymeyer et~al.,}{Nymeyer
  et~al.}{2011}]{Nymeyer_2011}
Nymeyer S.,  et~al., 2011, \mn@doi [\apj] {10.1088/0004-637X/742/1/35}, 742, 35

\bibitem[\protect\citeauthoryear{{Pierrehumbert}}{{Pierrehumbert}}{2010}]{Pierrehumbert_2010}
{Pierrehumbert} R.~T.,  2010, {Principles of Planetary Climate}

\bibitem[\protect\citeauthoryear{Puig, Pilbratt, Heske, Escudero~Sanz  \&
  Crouzet}{Puig et~al.}{2016}]{Puig_2016}
Puig L.,  Pilbratt G.~L.,  Heske A.,  Escudero~Sanz I.,   Crouzet P.-E.,  2016,
  in Space Telescopes and Instrumentation 2016: Optical, Infrared, and
  Millimeter Wave. p. 99041W, \mn@doi{10.1117/12.2230964}

\bibitem[\protect\citeauthoryear{Ranjan, Charbonneau, D{\'e}sert, Madhusudhan,
  Deming, Wilkins  \& Mandell}{Ranjan et~al.}{2014}]{Ranjan_2014}
Ranjan S.,  Charbonneau D.,  D{\'e}sert J.-M.,  Madhusudhan N.,  Deming D.,
  Wilkins A.,   Mandell A.~M.,  2014, \mn@doi [\apj]
  {10.1088/0004-637X/785/2/148}, 785, 148

\bibitem[\protect\citeauthoryear{Rasmussen \& Williams}{Rasmussen \&
  Williams}{2006}]{Rasmussen_2006}
Rasmussen C.,  Williams C.,  2006, Gaussian Processes for Machine Learning.
Adaptative computation and machine learning series, University Press Group
  Limited

\bibitem[\protect\citeauthoryear{Roman \& Rauscher}{Roman \&
  Rauscher}{2019}]{Roman_2018}
Roman M.,  Rauscher E.,  2019, \mn@doi [The Astrophysical Journal]
  {10.3847/1538-4357/aafdb5}, 872, 1

\bibitem[\protect\citeauthoryear{Rothman et~al.,}{Rothman
  et~al.}{2010}]{Rothman_2010}
Rothman L.~S.,  et~al., 2010, \mn@doi [\jqsrt] {10.1016/j.jqsrt.2010.05.001},
  111, 2139

\bibitem[\protect\citeauthoryear{Rybicki \& Lightman}{Rybicki \&
  Lightman}{2004}]{Rybicki_2004}
Rybicki G.~B.,  Lightman A.~P.,  2004, Radiative Processes in Astrophysics.
John Wiley \& Sons

\bibitem[\protect\citeauthoryear{Schwartz \& Cowan}{Schwartz \&
  Cowan}{2015}]{Schwartz_2015}
Schwartz J.~C.,  Cowan N.~B.,  2015, \mn@doi [\mnras] {10.1093/mnras/stv470},
  449, 4192

\bibitem[\protect\citeauthoryear{Schwarz}{Schwarz}{1978}]{Schwarz_1978}
Schwarz G.,  1978, \mn@doi [Ann. Statist.] {10.1214/aos/1176344136}, 6, 461

\bibitem[\protect\citeauthoryear{Seager, Richardson, Hansen, Menou, Cho  \&
  Deming}{Seager et~al.}{2005}]{Seager_2005}
Seager S.,  Richardson L.~J.,  Hansen B. M.~S.,  Menou K.,  Cho J. Y.-K.,
  Deming D.,  2005, \mn@doi [\apj] {10.1086/444411}, 632, 1122

\bibitem[\protect\citeauthoryear{Shporer et~al.,}{Shporer
  et~al.}{2014}]{Shporer_2014}
Shporer A.,  et~al., 2014, \mn@doi [\apj] {10.1088/0004-637X/788/1/92}, 788, 92

\bibitem[\protect\citeauthoryear{Stevenson et~al.,}{Stevenson
  et~al.}{2010}]{Stevenson_2010}
Stevenson K.~B.,  et~al., 2010, \mn@doi [\nat] {10.1038/nature09013}, 464, 1161

\bibitem[\protect\citeauthoryear{Stevenson et~al.,}{Stevenson
  et~al.}{2014a}]{Stevenson_2014}
Stevenson K.~B.,  et~al., 2014a, \mn@doi [Science] {10.1126/science.1256758},
  346, 838

\bibitem[\protect\citeauthoryear{Stevenson, Bean, Madhusudhan  \&
  Harrington}{Stevenson et~al.}{2014b}]{Stevenson_2014b}
Stevenson K.~B.,  Bean J.~L.,  Madhusudhan N.,   Harrington J.,  2014b, \mn@doi
  [\apj] {10.1088/0004-637X/791/1/36}, 791, 36

\bibitem[\protect\citeauthoryear{Stevenson et~al.,}{Stevenson
  et~al.}{2017}]{Stevenson_2017}
Stevenson K.~B.,  et~al., 2017, \mn@doi [\aj] {10.3847/1538-3881/153/2/68},
  153, 68

\bibitem[\protect\citeauthoryear{Swain et~al.,}{Swain
  et~al.}{2013}]{Swain_2013}
Swain M.,  et~al., 2013, \mn@doi [\icarus] {10.1016/j.icarus.2013.04.003}, 225,
  432

\bibitem[\protect\citeauthoryear{Todorov, Deming, Burrows  \&
  Grillmair}{Todorov et~al.}{2014}]{Todorov_2014}
Todorov K.~O.,  Deming D.,  Burrows A.,   Grillmair C.~J.,  2014, \mn@doi
  [\apj] {10.1088/0004-637X/796/2/100}, 796, 100

\bibitem[\protect\citeauthoryear{Waldmann, Tinetti, Rocchetto, Barton,
  Yurchenko  \& Tennyson}{Waldmann et~al.}{2015}]{Waldman_2015}
Waldmann I.~P.,  Tinetti G.,  Rocchetto M.,  Barton E.~J.,  Yurchenko S.~N.,
  Tennyson J.,  2015, \mn@doi [\apj] {10.1088/0004-637X/802/2/107}, 802, 107

\bibitem[\protect\citeauthoryear{Werner et~al.,}{Werner
  et~al.}{2004}]{Werner_2004}
Werner M.~W.,  et~al., 2004, \mn@doi [\apjs] {10.1086/422992}, 154, 1

\bibitem[\protect\citeauthoryear{Wilkins, Deming, Madhusudhan, Burrows,
  Knutson, McCullough  \& Ranjan}{Wilkins et~al.}{2014}]{Wilkins_2014}
Wilkins A.~N.,  Deming D.,  Madhusudhan N.,  Burrows A.,  Knutson H.,
  McCullough P.,   Ranjan S.,  2014, \mn@doi [\apj]
  {10.1088/0004-637X/783/2/113}, 783, 113

\bibitem[\protect\citeauthoryear{Wong et~al.,}{Wong et~al.}{2015}]{Wong_2015}
Wong I.,  et~al., 2015, \mn@doi [\apj] {10.1088/0004-637X/811/2/122}, 811, 122

\bibitem[\protect\citeauthoryear{Wong et~al.,}{Wong et~al.}{2016}]{Wong_2016}
Wong I.,  et~al., 2016, \mn@doi [\apj] {10.3847/0004-637X/823/2/122}, 823, 122

\bibitem[\protect\citeauthoryear{{Yurchenko} \& {Tennyson}}{{Yurchenko} \&
  {Tennyson}}{2014}]{YurchenkoTennyson2014mnrasExomolCH4}
{Yurchenko} S.~N.,  {Tennyson} J.,  2014, \mn@doi [\mnras]
  {10.1093/mnras/stu326}, \href
  {http://adsabs.harvard.edu/abs/2014MNRAS.440.1649Y} {440, 1649}

\bibitem[\protect\citeauthoryear{Zhao, Milburn, Barman, Hinkley, Swain, Wright
  \& Monnier}{Zhao et~al.}{2012}]{Zhao_2012}
Zhao M.,  Milburn J.,  Barman T.,  Hinkley S.,  Swain M.~R.,  Wright J.,
  Monnier J.~D.,  2012, \mn@doi [\apjl] {10.1088/2041-8205/748/1/L8}, 748, L8

\bibitem[\protect\citeauthoryear{Zhou, Bayliss, Kedziora-Chudczer, Tinney,
  Bailey, Salter  \& Rodriguez}{Zhou et~al.}{2015}]{Zhou_2015}
Zhou G.,  Bayliss D. D.~R.,  Kedziora-Chudczer L.,  Tinney C.~G.,  Bailey J.,
  Salter G.,   Rodriguez J.,  2015, \mn@doi [\mnras] {10.1093/mnras/stv2138},
  454, 3002

\bibitem[\protect\citeauthoryear{Zingales \& Waldmann}{Zingales \&
  Waldmann}{2018}]{Zingales_2018}
Zingales T.,  Waldmann I.~P.,  2018, \mn@doi [The Astronomical Journal]
  {10.3847/1538-3881/aae77c}, 156, 268

\makeatother
\end{thebibliography}


\bsp	
\label{lastpage}
\end{document}